\documentclass[fleqn,usenatbib]{mnras}

\usepackage{newtxtext,newtxmath}

\usepackage[T1]{fontenc}

\DeclareRobustCommand{\VAN}[3]{#2}
\let\VANthebibliography\thebibliography
\def\thebibliography{\DeclareRobustCommand{\VAN}[3]{##3}\VANthebibliography}

\usepackage{graphicx}	
\usepackage{amsmath}	
\usepackage{pdflscape}
\usepackage{multirow}
\usepackage[table,xcdraw]{xcolor}

    \title[GRS 1915+105 in the `obscured' phase]{Multi-mission view of low-luminosity `obscured' phase of GRS 1915+105}

\author[Athulya et al.]{Athulya M. P.,$^{1, 3}$\thanks{E-mail: athulyamp9425@gmail.com, a.madathil-pottayil@herts.ac.uk, \newline Affiliation with Dayananda Sagar University ended on 15$^{\rm th}$ of September 2022.}
Anuj Nandi$^{2}$
\\
$^{1}$Department of Physics, Dayananda Sagar University, Bengaluru, 560068, India\\
$^{2}$Space Astronomy Group, ISITE Campus, U R Rao Satellite Centre, Bengaluru, 560037, India \\
$^{3}$Centre for Astrophysics Research, University of Hertfordshire, College Lane, Hatfield, AL10 9AB, UK
\\
}

\date{Accepted XXX. Received YYY; in original form ZZZ}

\pubyear{2022}

\begin{document}
\label{firstpage}
\pagerange{\pageref{firstpage}--\pageref{lastpage}}
\maketitle

\begin{abstract}
GRS~1915+105 is observed in an `obscured' phase since May 2019, exhibiting  steady and low X-ray luminosities, while being intervened by sporadic re-brightenings. In this work, we perform a comprehensive and wide-band analysis of the spectral and timing properties of the source during the period 2019--2021 using \textit{AstroSat} (\textit{SXT}: 0.5--8 keV; \textit{LAXPC}: 3--60 keV), \textit{NICER} (0.5--12 keV), and  \textit{NuSTAR} (3--60 keV) observations. Spectral analysis reveals the presence of a highly variable obscurer (N$_{\rm H_{\rm 1}}\sim$ $10^{22}$--$10^{24}$ atoms cm$^{-2}$) throughout the observation period. Source is detected in the Low/Hard state for most of the time, with the spectra being described by a Comptonised component ($\Gamma$~$\sim$1.16 -- 1.79, kT$_{\rm e}\sim$2 -- 31 keV).  The source spectra steepen ($\Gamma\sim2.5$) indicating softening of the spectrum during the rise of the re-brightenings.  {Various emission and absorption lines corresponding to the neutral Fe-K$\alpha$, Fe-XXV K$\alpha$, Fe-XXVI K$\alpha$, and the Ni-XXVIII K$\alpha$ were detected with the equivalent widths varying between 70 eV -- 3.5 keV.} The column density of the absorbing plasma varied between $10^{16}$--$10^{18}$ atoms cm$^{-2}$ at a distance $\leq2\times$10$^{10}$ cm. Interestingly, the source is also seen exhibiting various variability classes ($\rho, \lambda, \delta, \chi$) 
at relatively low luminosities ($\sim$0.01L$_\textrm{Edd}$) during the re-brightening phases. Different variability classes show signature of QPOs ($\nu_{\rm QPO}$: 20--180 mHz, 
rms$_{\rm QPO}$: 7.5\%--16\%). The source showed a maximum bolometric luminosity {(L$_\textrm{bol}$)} of $\sim$0.01L$_\textrm{Edd}$ (Re-brightening phases) and a minimum L$_\textrm{bol}$ of 0.004L$_\textrm{Edd}$ (Quiet phase) during the period. We discuss the possible disc dynamics around the black hole during this low-luminosity `obscured' phase.
\end{abstract}




\section{Introduction}
\label{sec1}

GRS 1915+105 is a unique Low Mass X-ray Binary (LMXB), hosting a massive (12.4$_{-1.8}^{+2.0}$M$_\odot$; \citealt{2014ApJ...796....2R}) and a maximally rotating ($\hat{a} > 0.98^{+0.01}_{-0.01}$; \citealt{2006ApJ...652..518M, 2009ApJ...706...60B, 2020MNRAS.499.5891S}) black hole at the center, accreting matter from a K-giant companion \citep{2001A&A...373L..37G}. GRS 1915+105 is the only LMXB that has exhibited 15 unique variability classes  ($\alpha, \beta, \gamma, \delta, \theta, \kappa, \lambda, \mu, \nu, \rho, \phi, \chi, \omega, \eta, \xi$) so far (\cite{2000A&A...355..271B, 2002MNRAS.331..745K, 2005A&A...435..995H}, see also \citealt{2022MNRAS.510.3019A, 2022MNRAS.512.2508M}). Each of these classes exhibit variabilities at timescales ranging from a few seconds to many hours, thereby providing a captivating illustration of various instabilities and the timescales at which these instabilities develop in the accretion disc around a stellar mass black hole (\citealt{2000ApJ...542L..33J, 2001MNRAS.324..267N, 2001A&A...380..245N, 2004ApJ...607..410T, 2012ApJ...750...71N} and the references therein). GRS~1915+105 exhibits steady radio jets in the hard state \citep{2000A&A...362..691N, 2002MNRAS.331..745K}, while transient and discrete radio jets \citep{1999MNRAS.304..865F, 1999ApJ...511..398R} are seen during transition of the source  from the hard state to the soft state. Besides jet events, disc winds have also been a prominent ejection event in GRS 1915+105. \cite{2000ApJ...539..413K, 2002ApJ...567.1102L, 2009ApJ...695..888U, 2016ApJ...833..165Z, 2018ApJ...860L..19N, 2022arXiv220302659L} detected various absorption features that indicated ionized outflowing winds, which also acted as jet suppressing mechanisms by averting the disc matter inflow into the radio jets \citep{2008ApJ...680.1359M, 2009Natur.458..481N, 2011AAS...21740401N}. Owing to its eccentric inflow and outflow phenomena, GRS 1915+105 sets an exemplary case study to analyse the astrophysical phenomena around the compact object.

After more than 25 years of high X-ray activity, GRS 1915+105 began to show a decrease in the X-ray flux \citep{2018ATel11828....1N, 2022MNRAS.510.3019A} since May 2018, leading to the presumption that the source is finally approaching quiescence. Yet, exceptionally again, the source started exhibiting a series of non-generic activities since April 2019 \citep{2019ATel12742....1H}. The detection of multiple absorption lines in the energy spectrum of GRS 1915+105 \citep{2020ApJ...904...30M} and the requirement of an additional absorption model to address the  soft excess \citep{2021ApJ...909...41B}, indicated the presence of a local obscuration in the system. \citealt{2022MNRAS.510.3019A} also detected a decrease in the bolometric luminosity of the source, caused by the local obscuration in the system. GRS 1915+105 is therefore, perceived to have entered into a new state of accretion called the `obscured state'. Obscuration, although is a commonly observed phenomenon in Active Galactic Nuclei (\citealt{1984MNRAS.208...15S, 2002ApJ...573..505Z, 2018KPCB...34..302V} and the references therein), is rarely observed in LMXBs. The various models developed to explain the cause of obscuration in X-ray binaries comprise the disc flaring theory \citep{1978A&A....67L..25M}, the obscuring winds caused by the stellar activity in the secondary star \citep{1982ApJ...253L..61W, 2014ApJ...786L..20M}, slim disc at the close vicinity of the compact object \citep{2017MNRAS.471.1797M} etc. However, in the recent work on GRS 1915+105, \citealt{2020ApJ...902..152N} detected three ionization zones layered up at a distance $\sim10^{11}$ cm, around the outer disc. Their study featured the possibility of vertical expansion of the outer disc that further acted as the obscuration medium. Meanwhile, \citealt{2020ApJ...904...30M} estimated the wind launch radius (r < $\sim10^{9}$ cm), velocity of the winds (350 km s$^{-1}$), and the magnetic field strength required to drive the winds away from the compact object. Their results revealed a wind that failed to escape the system, eventually enshrouding the compact object thus causing the obscuration.

Over the period of obscuration, the source also displayed several re-brightenings \citep{2019ATel12761....1I, 2020ATel13974....1R, 2021ATel14792....1N}, either in the form of a quick flare or a prolonged re-brightening. Few quick flares were reported to be a sequel to the radio flares  \citep{2019ATel13304....1T, 2019ATel12773....1M, 2020ATel13442....1T, 2020ATel13652....1A}, whereas during the prolonged re-brightenings the \textit{ALMA} observations (15.5 GHz) of the source showed a decrease in the radio activity (\citealt{2021MNRAS.503..152M}). The quasi-simultaneous radio-X-ray flares happening on short timescales ($\sim1400$ sec) was also observed in Cyg X--1 \citep{2005Natur.436..819G, 2006MNRAS.369..603F, 2007ApJ...663L..97W}, where the X-ray emission is hypothesised to be originated at the base of the jet. The prolonged re-brightenings (also called as re-flares, mini-outbursts, failed outbursts etc.) have been observed in many LMXBs. The nature of the mini-outbursts varied in each LMXBs, with a few sources exhibiting only one spectral state throughout the mini-outburst (eg: MAXI J1659--152 \citep{2013ApJ...775....9H}, IGR J17379--3747 \citep{2019ApJ...877...70B}, XTE J1650--500 \citep{2004ApJ...601..439T, 2017MNRAS.470.4298Y}, while few other sources exhibited different spectral states throughout the mini-outburst (MAXI J1535--571 \citep{2020MNRAS.496.1001C} and GRS 1739--278 \citep{2017MNRAS.470.4298Y}. Irrespective of the nature of the outbursts, the cause and offset of the mini-outbursts are not clearly understood. Augmented mass transfer due to the irradiation of the companion \citep{1993ApJ...408L...5C, 1993A&A...279L..13A, 2000A&A...353..244H} is one of the commonly used models to explain the cause of a mini-outburst.

GRS~1915+105 has been extensively studied throughout 26-years of long outburst. However, only a few attempts, using scattered observations of the source, have been made to understand the characteristics after the source descended into the low-luminosity `obscured' phase. In this manuscript, we perform for the first time, an in-depth and a cohesive analysis of the spectral and timing properties of the source, during the period of March 2019 to  {November 2021} using observations from \textit{AstroSat}, \textit{NICER} and \textit{NuSTAR}. Through our results, we describe the attributes of obscuration in the system. The observations also reveal the source to be exhibiting multiple re-brightening phases with the display of the characteristic variability classes and the transition between classes during prolonged re-brightenings. We, therefore, characterize the source properties and spectral state transitions observed during the re-brightening phases as well as the quiet phase, in this work.

This paper is structured as follows: \S \ref{sec 2} briefly describes the data reduction procedures for all the observations obtained from \textit{SXT \& LAXPC} onboard \textit{AstroSat}, \textit{NICER} and \textit{NuSTAR}. In \S \ref{sec 3}, we explain the modeling techniques and the procedure of spectral and timing analysis. In \S \ref{sec 4}, we present the results obtained through our analysis and in \S \ref{sec 5}, we discuss the overall behavioural pattern of the source. Finally, in \S \ref{sec 6}, we conclude with a summary of our results.

\begin{figure*}
  \centering
  \hspace*{-1cm}
  \begin{minipage}[b]{0.35\textwidth}
   \hspace*{1cm}
    \includegraphics[width=8cm, height=5.2cm]{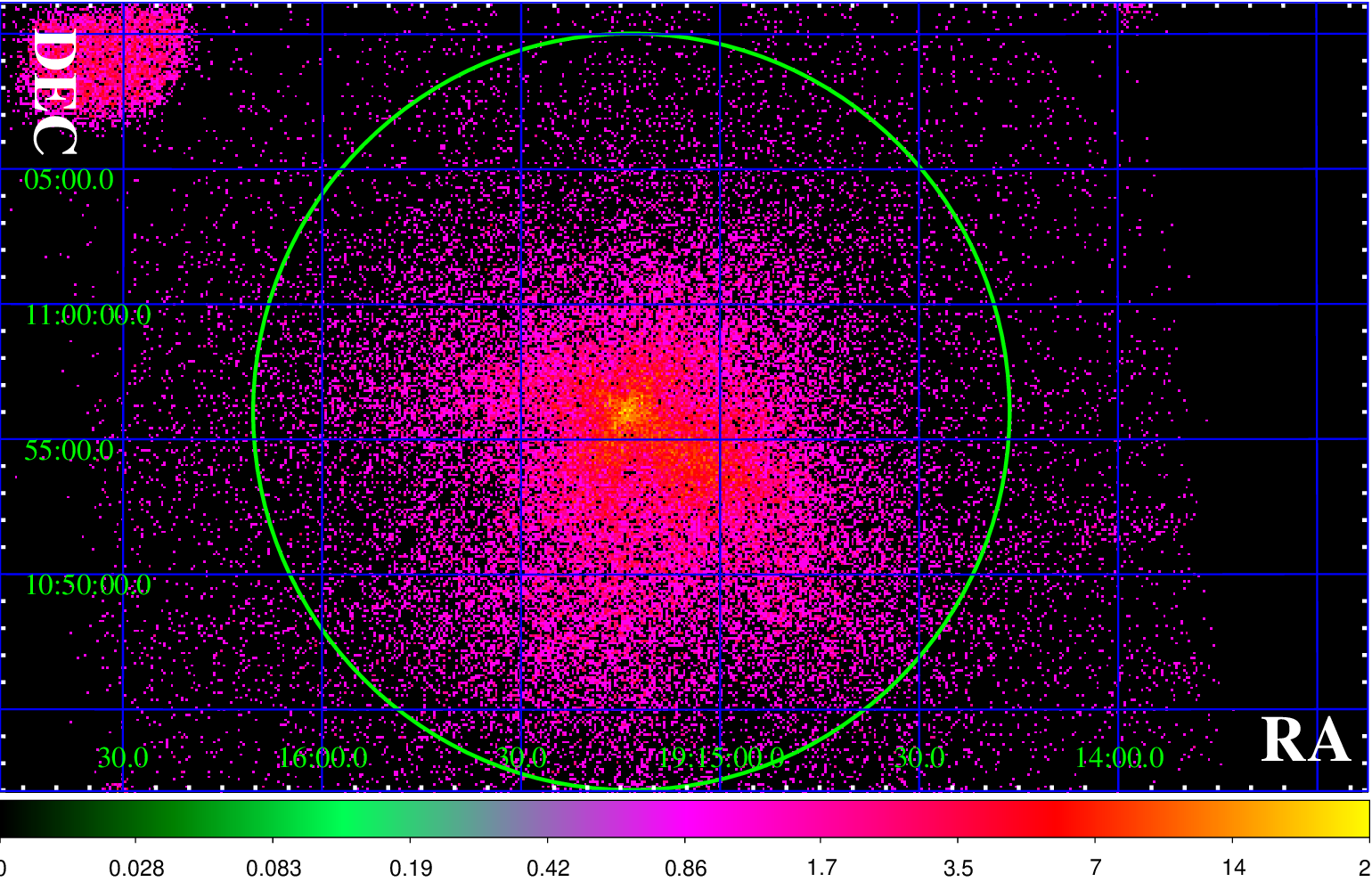}
     \label{fig 1a}
  \end{minipage}
  \hfill
  \begin{minipage}[b]{0.35\textwidth}
  \hspace*{-1.8cm}
    \includegraphics[width=8cm,height=5.2cm]{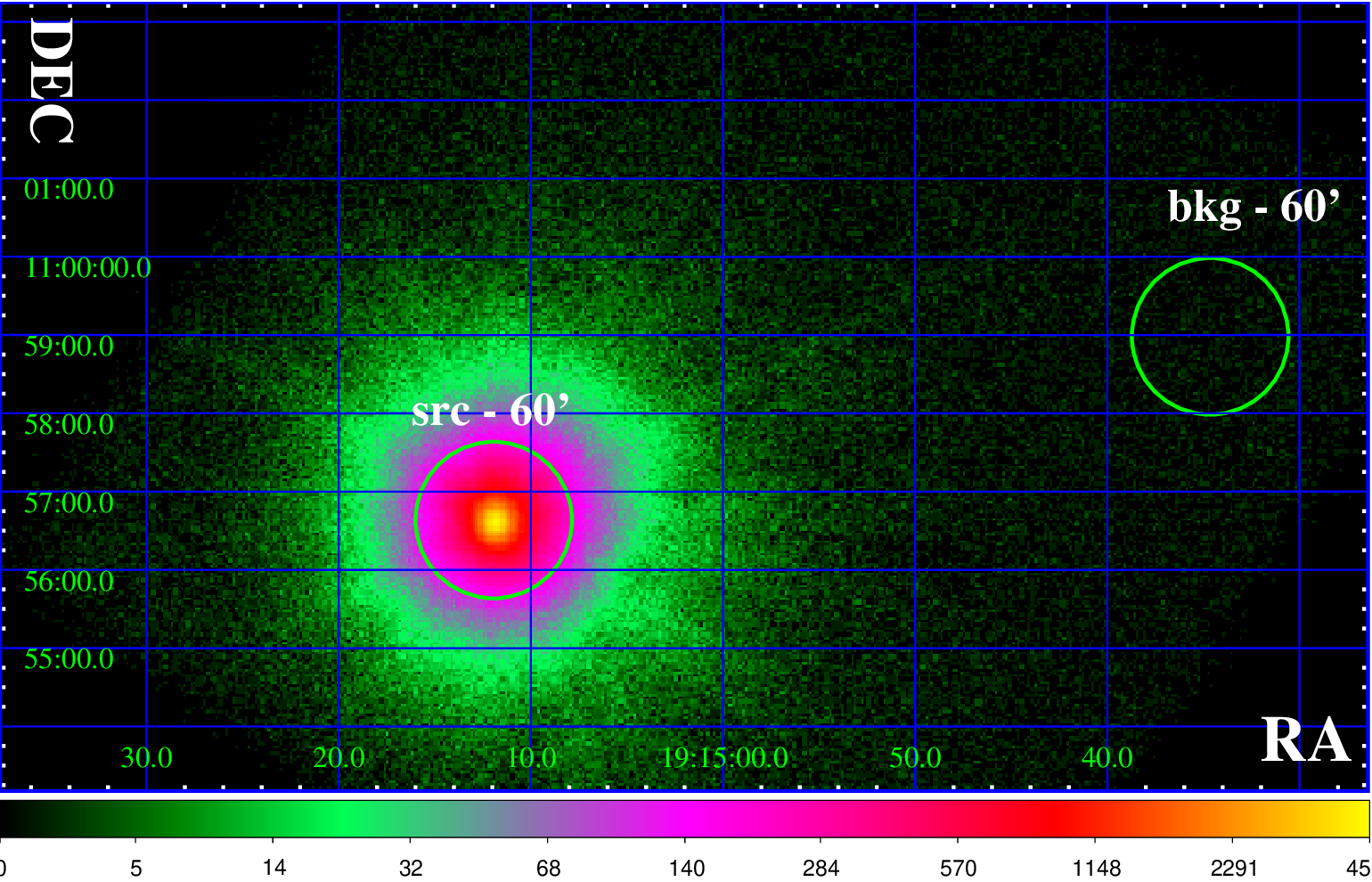}
    \label{fig 1b}
  \end{minipage}
  \caption{\textit{Left:} The image of GRS 1915+105, taken in the PC mode, on MJD 59123 (Epoch AST10) by \textit{AstroSat-SXT}. The green circle of 12$^{\prime}$ radius encloses the region selected for the extraction of  the source products. \textit{Right:}  GRS 1915+105 as observed by \textit{NuSTAR-FPMA}  during MJD 59409 (Epoch NuS13). Two green circles of 60$^{\prime\prime}$ radius, { one encircling the source and the other source-free region, show the extraction regions for both source and background products}. \label{fig 1}}
\end{figure*}

\section{Observations and Data Reduction}
\label{sec 2}

We use the data obtained from \textit{AstroSat, NICER} and \textit{NuSTAR} from March 2019 to  {November 2021} to  perform a coordinated and wide-band study of the spectral and timing properties of the source.  {Table \ref{tab 1} enlists the overall observations of the source made by \textit{AstroSat} together with the simultaneous \textit{NICER -- NuSTAR} observations available during these Epochs. All of them are also indicated in Figure \ref{fig 2A} with vertical dashed lines. In addition, Table \ref{tab 2} gives the log of further \textit{NICER} observations for most of the re-brightening phases observed between March 2019 –-  {November 2021}.} Below, we brief the reduction procedures for the data obtained from all three instruments.

\subsection{\textit{AstroSat}}
\label{sec 2.1}

\textit{AstroSat} \citep{2006AdSpR..38.2989A}, India's first dedicated astronomy mission, observes celestial bodies in broad energy band simultaneously, ranging from near UV to hard  X-rays. In our work, we use the observations made by two instruments on board \textit{AstroSat}, the Soft X-ray Telescope (\textit{SXT}) \citep{2016SPIE.9905E..1ES, 2017JApA...38...29S} covering the energy band $0.3 - 8$ keV and the Large Area X-ray Proportional Counter (\textit{LAXPC}) \citep{2016ApJ...833...27Y, 2017ApJS..231...10A} operating in $3 - 80$ keV energy band. \textit{AstroSat} had made 5 observations of the source during our period of study. Level-2 \textit{SXT} data and Level-1 \textit{LAXPC}20 data are obtained from the ISSDC data dissemination archive\footnote{\url{https://astrobrowse.issdc.gov.in/astro_archive/archive/Home.jsp}}. In order to perform simultaneous analysis, we choose one segment from both \textit{SXT} and \textit{LAXPC}, where the observations made by both these instruments happen at almost same time, with exposure time $\geq$ 1 ks for both  instruments.  {The \textit{SXT} pipeline \texttt{AS1SXTLevel2-1.4b} is used for Level 2 data analysis. The light curve and the spectral files are further extracted using the \texttt{XSELECT V2.5a}. The background spectrum and the response files for \textit{SXT} data is  distributed by TIFR-POC\footnote{\url{https://www.tifr.res.in/~astrosat_sxt/sxtpipeline.html}}. The ARF files for each observation is generated using the \texttt{sxtARFModule} tool. The \textit{LAXPC} data processing, from Level 1 to Level 2, is carried out using the software \texttt{LaxpcSoft v 3.4}. The background spectrum for each \textit{LAXPC} spectrum is generated using the code \texttt{backshiftv3.e}, while we use the pre-computed response files (version v1.0) provided by the \textit{LAXPC} team\footnote{\url{https://www.tifr.res.in/~astrosat_laxpc/LaxpcSoft.html}}. A detailed description for the standard reduction and extraction procedures are provided in \citealt{2022MNRAS.510.3019A} (see also \citealt{2019MNRAS.487..928S, 2020MNRAS.497.1197B}).} Additionally, a  circular region of 12$^\prime$ is used while extracting \textit{SXT} data (see Figure \ref{fig 1}). We use a combination of the top-layer and all events during the extraction of \textit{LAXPC} data. The energy spectra thus obtained are grouped to 25 counts per bin using the \texttt{FTOOL - grppha} and a systematic error of 3\% \citep{2021JApA...42...32A} is additionally included to account for the uncertainty in the spectral response.

\subsection{\textit{NuSTAR}}
\label{sec 2.2}
The Nuclear Spectroscopic Telescope Array (\textit{NuSTAR}) is sensitive to X-rays within the energy range $3 - 78$ keV \citep{2013ApJ...770..103H}. In this paper, we consider all the 11 observations of the source made by the modules, FPMA and FPMB onboard \textit{NuSTAR}, during the period 2019 -- 2021. These data were obtained from the HEASARC database\footnote{\url{https://heasarc.gsfc.nasa.gov/cgi-bin/W3Browse/w3browse.pl}}. \textit{NuSTAR} Data Analysis Software (NuSTARDAS) \texttt{nupipeline V 0.4.9} and CALDB v20191219 is used to generate the cleaned event file \citep{2022arXiv220413363P}. A circular region of 60$^{\prime\prime}$ is used to extract the source events. Similarly, a region of  60$^{\prime\prime}$ radius that is free of source photons, is chosen for the background extraction (see Figure \ref{fig 1}). The cleaned events and the region files thus obtained are used to extract the source products using the module \texttt{nuproducts}. The extracted spectra were uniformly grouped to 25 counts per energy bin and is modeled in the energy range 3 -- 60 keV.

\subsection{\textit{NICER}}
\label{sec 2.3}

 {Neutron star Interior Composition Explorer (\textit{NICER}) \citep{2012SPIE.8443E..13G} has persistently observed GRS 1915+105 during the obscured phase using its primary scientific instrument, the X-ray Timing Instrument (\textit{XTI}) that covers the energy band $0.2 - 12$ keV. In our work, we study  {23} \textit{NICER-XTI} observations, some of which without a simultaneous high energy observation.  {We also did a thorough spectral analysis of 40 additional observations, made by \textit{NICER} between MJD $58610-58650$, $59050-59150$ and $59375-59500$. However, we tabulate only 23 observations, as they sufficiently describe the spectral and timing evolution of the source parameters. }These observations were obtained from  the HEASARC database\footnote{\url{https://heasarc.gsfc.nasa.gov/cgi-bin/W3Browse/w3browse.pl}}. The data is processed using the latest version of the \textit{NICER} software (NICERDAS ver 10). The Level 2 analysis is performed using the \texttt{nicerl2} task. The Level 3 data analysis is performed using the new extraction task \texttt{nicerl3-spect} (recently made available with \texttt{HEASoft 6.31} released in November 2022). \texttt{nicerl3-spect}  concurrently generates spectrum, background, ancillary and response files from the un-barycentered merged event file. Additionally, we also set the background model type to 3C50 while running the \texttt{nicerl3-spect} script. All the spectra thus obtained were uniformly grouped to 25 counts per bin.}

\begin{table*}
\centering
\caption{Log of \textit{AstroSat, NICER} and \textit{NuSTAR} observations of GRS~1915$+$105 from March 2019 to  {November 2021}.
 The table enlists the  details of the every observation: The Epoch of observation, Mission ID, Observation ID (Obs. ID), Date of observation, instrument exposure time, source count rate, events, and variability class categorization during each observation.  {The rows shaded in grey represent the individual \textit{NuSTAR} observations, whereas the unshaded rows includes the simultaneous observations from the instruments \textit{SXT+LAXPC} and \textit{NICER+NuSTAR}.}}
\begin{tabular}{|lcccccccc|}
\hline
Epoch & Mission & Observation & Date  & Inst. & Exp. time & Avg. Cts$\dagger$ & Events & Variability \\
& (ID) & (Obs. ID) & UTC (MJD) & & (sec)  & (cts s$^{-1}$) & &  Class\\
\hline
                & & & &  \textit{SXT}   & 1110 & 6 & &\\
\multirow{-2}{*}{1} & \multirow{-2}{*}{AST1}  & \multirow{-2}{*}{A05\_173T01\_9000002812}  &  \multirow{-2}{*}{23/3/2019 (58565)} & \textit{LAXPC}  & 3000 & 300 & \multirow{-2}{*}{decay} & \multirow{-2}{*}{$\chi$}\\
\hline
\rowcolor{lightgray}  2 & NuS2 & 90501321002              &  5/5/2019 (58608) &  \textit{NuSTAR}   & 3250 & 31& decay & - \\
\hline
                  &  &  & &   \textit{SXT}   & 910 & 2 & &\\
\multirow{-2}{*}{3} & \multirow{-2}{*}{AST3}  & \multirow{-2}{*}{T03\_116T01\_9000002916} & \multirow{-2}{*}{15/5/2019 (58618)} & \textit{LAXPC} & 2540 & 97 & \multirow{-2}{*}{RB$_{\rm I}$} & \multirow{-2}{*}{-}\\

\hline
                   & NIC4 & 2596011202 & &  \textit{NICER}   & 1015 & 7 & &\\
\multirow{-2}{*}{4}  & NuS4 & 30502008002         & \multirow{-2}{*}{19/5/2019 (58622)} &  \textit{NuSTAR}   & 3235 & 7 & \multirow{-2}{*}{ {RB$_{\rm I}$/Radio Flare}} & \multirow{-2}{*}{-}\\
\hline
                   &  &  & &  \textit{SXT}   & 2200 & 1 & & \\
\multirow{-2}{*}{5}  & \multirow{-2}{*}{AST5}  & \multirow{-2}{*}{T03\_117T01\_9000002988 }  & \multirow{-2}{*} {14/6/2019 (58648)} & \textit{LAXPC} & 2840 & 93 & \multirow{-2}{*}{ {RB$_{\rm I/}$/Decay}} & \multirow{-2}{*}{-}   \\
\hline
                   & NIC6 & 2596012201       & & \textit{NICER}   & 580 & 6 & & \\
\multirow{-2}{*}{6} & NuS6 &  30502008004      & \multirow{-2}{*}{31/7/2019 (58695)} &\textit{NuSTAR}   & 3270 & 14 & \multirow{-2}{*}{Quiet} & \multirow{-2}{*}{$\chi$} \\
\hline
                   & NIC7 &2596012804               &  &\textit{NICER}   & 997 & 10 &  &  \\
\multirow{-2}{*}{7} & NuS7 & 30502008006             & \multirow{-2}{*}{13/9/2019 (58739)} &  \textit{NuSTAR}   & 3095 & 24 & \multirow{-2}{*}{Quiet} & \multirow{-2}{*}{$\chi$} \\
\hline
\rowcolor{lightgray}  8 & NuS8 & 90501346002              & 16/10/2019 (58772) & \textit{NuSTAR}   & 3235 & 11 &Quiet& $\chi$ \\
\hline
                    & NIC9 & 3607010701               &  & \textit{NICER}   & 2530 & 4 & &  \\
\multirow{-2}{*}{9} & NuS9 & 30602008002              & \multirow{-2}{*}{15/4/2020 (58954)} & \textit{NuSTAR}   & 2880 & 5 & \multirow{-2}{*}{RB$_{\rm V}$} & \multirow{-2}{*}{$\chi$} \\
\hline
                    &  &  &  &  \textit{SXT}   & 1620 & 4 &  &  \\
\multirow{-2}{*}{10} & \multirow{-2}{*}{AST10}  & \multirow{-2}{*}{T03\_231T01\_9000003910}  & \multirow{-2}{*}{1/10/2020 (59123)} & \textit{LAXPC} & 2490 & 201 & \multirow{-2}{*}{RB$_{\rm V}$} & \multirow{-2}{*}{$\rho$}\\
\hline
\rowcolor{lightgray}  11 & NuS11 & 30602008004   & 21/10/2020 (59143) &  \textit{NuSTAR}   & 3300 & 3 & RB$_{\rm V}$ & $\chi$\\
\hline
\rowcolor{lightgray}  12 & NuS12 & 30602008006   & 29/11/2020 (59182) & \textit{NuSTAR}   & 3230 & 27 & Quiet & $\chi$\\
\hline
                    & NIC13 & 4647012002               &  &  \textit{NICER}   & 2230 & 205 & &\\
\multirow{-2}{*}{13}& NuS13 & 90701323002              & \multirow{-2}{*}{14/7/2021 (59409)} &  \textit{NuSTAR}   & 3180 & 76 & \multirow{-2}{*}{RB$_{\rm VI}$} & \multirow{-2}{*}{$\delta$} \\
\hline
                      &  &  &  & \textit{SXT}   & 13970 & 5 &  &\\
\multirow{-2}{*}{14}  & \multirow{-2}{*}{AST14}  & \multirow{-2}{*}{T04\_029T01\_9000004582}  & \multirow{-2}{*}{22/7/2021 (59417)} &   \textit{LAXPC} & 16890 & 312 & \multirow{-2}{*}{RB$_{\rm VI}$} & \multirow{-2}{*}{$\delta$}\\
\hline
                     &  &  &  &  \textit{SXT}   & 82840 & 5 &  &\\
\multirow{-2}{*}{15} & \multirow{-2}{*}{AST15} & \multirow{-2}{*}{T04\_029T01\_9000004582} & \multirow{-2}{*}{23/7/2021 (59418)} &  \textit{LAXPC}   & 83300 & 277 &  \multirow{-2}{*}{RB$_{\rm VI}$} & \multirow{-2}{*}{$\delta$}\\
\hline
                     &  &  &  & \textit{SXT}   & 35260 & 7 &  &  \\
\multirow{-2}{*}{16} & \multirow{-2}{*}{AST16}   & \multirow{-2}{*}{T04\_029T01\_9000004582}  & \multirow{-2}{*}{24/7/2021 (59419)} &   \textit{LAXPC} & 54770 & 327 & \multirow{-2}{*}{RB$_{\rm VI}$} & \multirow{-2}{*}{$\delta$}\\
\hline
                    & NIC17 & 4103010270   &  &  \textit{NICER}    & 1210 & 12 & & \\       
\multirow{-2}{*}{17} &NuS17 & 90701332002  & \multirow{-2}{*}{30/9/2021 (59487)} &  \textit{NuSTAR}   & 3040 & 5 & \multirow{-2}{*}{RB$_{\rm VI}$} & \multirow{-2}{*}{$\chi$} \\
\hline
                     & NIC18 & 4647013801   &  &  \textit{NICER}   & 1330 & 65 &  &\\
\multirow{-2}{*}{18} & NuS18 & 90701335002  & \multirow{-2}{*}{15/11/2021 (59533)} & \textit{NuSTAR}  & 3115 & 4 & \multirow{-2}{*}{Quiet} & \multirow{-2}{*}{$\chi$} \\

\hline
\end{tabular}
\begin{flushleft}
      $\dagger$ Average counts calculated for \textit{SXT} (0.7 -- 7 keV), \textit{LAXPC} (3 -- 60 keV), \textit{NICER} (0.7 -- 12 keV) and \textit{NuSTAR} (3 -- 60 keV) in the respective energy bands, mentioned in the parenthesis.
\end{flushleft}
\label{tab 1}
\end{table*}

\section{Analysis \& Modeling}
\label{sec 3}
\subsection{Timing Analysis}
\label{sec 3.1}

Light curves for \textit{AstroSat-LAXPC} and \textit{NICER} observations were initially generated with a time bin of 10 sec in the energy ranges 3 -- 60 keV and 0.5 -- 12 keV respectively. In addition, the \textit{NICER} light curves corresponding to each re-brightening phase observation were separately extracted again for three individual energy ranges; 0.3 -- 3 keV, 3 -- 6 keV and 6 -- 12 keV with a time bin of 10 sec in order to plot the Color-Color Diagram (CCD) (see Figures \ref{fig 3}, \ref{fig 4} and \ref{fig 5}). The CCD plots HR1 (Hardness Ratio 1) on the X - axis and HR2 (Hardness Ratio 2) on the Y - axis, where HR1 is the ratio of photons from 3 -- 6 keV to the photons in 0.3 -- 3 keV and  HR2 is the ratio of photons in 6 -- 12 keV to the photons in 0.3 -- 3 keV.

The Power Density Spectrum (PDS) was generated using light curves with a time resolution of 10 msec, obtained from  \textit{AstroSat-LAXPC, NICER} and  \textit{NuSTAR} observations. Although a Nyquist frequency of 50 Hz was obtained, the PDS was dominated with noise above 1 Hz. The data points were binned to 32768 bins resulting in a lowest frequency of 0.003 Hz (1/(32768×10 ms)). All the PDS are rms-normalized \citep{1990A&A...230..103B, 1999ApJ...510..874N} with a geometrical re-bin factor of 1.05. The noise associated with the PDS is fitted using \textit{powerlaw}. The narrow features in the PDS, known as Quasi-periodic Oscillations (QPOs), are described using the \textit{Lorentzian} profile,  
\begin{math}
    L(f) = \frac{KQf_{0}/\pi}{f_{0}^{2} + Q^{2}(f-f_{0})^{2}},
\end{math} where, $f_{0}$ is the frequency of the QPO,  {K is the normalization that defines the total strength of the oscillation and Q (= $f_{0}/\Delta$, where $\Delta$ is the half width at half maximum) is the quality factor that denotes the coherence of the variation.} 
We, therefore, use a \textit{Lorentzian} model to fit the QPO feature (\citealt{2012MNRAS.427..595M} and the references therein). Details of the method to obtain the model fitted parameters are mentioned in \citealt{2022MNRAS.510.3019A}. The narrow features with a  {Q-factor} $\geq$ 2 \citep{2000ARA&A..38..717V, 2006ARA&A..44...49R} and  {significance} ($\sigma$) $> 3$ \citep{2012MNRAS.426.1701B, 2019MNRAS.487..928S}  {are considered as QPOs}.  {The total rms variability for the frequency range 0.003 -- 1 Hz is estimated using the rectangle rule integration method, where the \textit{rms} = ${\sqrt{(P(\nu)\times\delta\nu)}}\times100$ (in \%). P($\nu$) is in units of rms$^2$ Hz$^{-1}$ and $\delta\nu$ is the interval width in Hz (see \citealt{2022MNRAS.510.3019A} and the references therein).}

\begin{table}
\centering 
\caption{Log of \textit{NICER} observations of GRS~1915$+$105 corresponding to the re-brightening phases between March 2019 --  {November 2021}. The Epoch of observation, source count rate and variability classes  are also included along with exposure time.}
\hspace*{-0.1cm}
\begin{tabular}{|lcccccc|}
\hline
\multicolumn{7}{|c|}{Re-brightening Phases} \\
\hline
Eve. & Epoch & Obs\_ID & Date & Avg Cts &  Exp. & Var. \\
& & & (MJD) & (cts  s$^{-1}$) & (sec) & Class \\
\hline
\multirow{6}{*}{RB$_{\rm I}$} & Obs. 1 &  2596011202   & 58622 & 7    & 1890 & -\\
                              & Obs. 2 &  2596011203   & 	58623.0 & 19   & 2415 & -\\
                              & Obs. 3 &  2596011203   & 	58623.9 & 120  & 810 & -\\
                              & Obs. 4 &  2596011206   & 58626 &  40  & 570 & -\\
                              & Obs. 5 &  2596011301   & 58631 &  2   & 830 & -\\
 \hline
\multirow{1}{*}{RB$_{\rm II}$} & Obs. 1 & 2103010204  & 58807 & 56 & 875 & -\\
\hline
\multirow{1}{*}{RB$_{\rm III}\dagger$}& - & - & - & - & - & -\\
\hline
\multirow{1}{*}{RB$_{\rm IV}$} & Obs. 1 &  3103010224 & 58995  & 428 &2618 & -\\
\hline
\multirow{6}{*}{RB$_{\rm V}$} & Obs. 1 & 3607012202    & 59063 & 17  & 1015 & $\chi$\\
                              & Obs. 2 & 3607012602   & 59088 & 105 & 995 & $\rho^\prime$\\
                              & Obs. 3 &  3607012901  & 59109 & 118 & 1060 & $\lambda$\\
                              & Obs. 4 &  3607013102  & 59127 & 28  & 1010 & $\rho$\\
\hline
\multirow{6}{*}{RB$_{\rm VI}$}  & Obs. 1 &  4647011801   & 59396 & 29    & 775  & $\chi$\\
                                & Obs. 2 &  4103010223   & 59405 & 203   & 1087 & $\delta$\\
                                & Obs. 3 &  4103010229   & 59413 & 256   & 990 & $\delta$\\
                                & Obs. 4 &  4103010237   & 59422 & 216   & 1040 & $\delta$\\
                                & Obs. 5 &  4103010271   & 59489 & 6     & 1180 & $\chi$\\

\hline
\end{tabular}
\begin{flushleft}
       $\dagger$Spectral and timing analysis of the source during  RB$_{\rm III}$ is not performed in this work due to the lack of observational data.
\end{flushleft}
\label{tab 2}
\end{table}

\begin{figure*}
\centering
\hspace*{-1cm}
\includegraphics[scale=0.58]{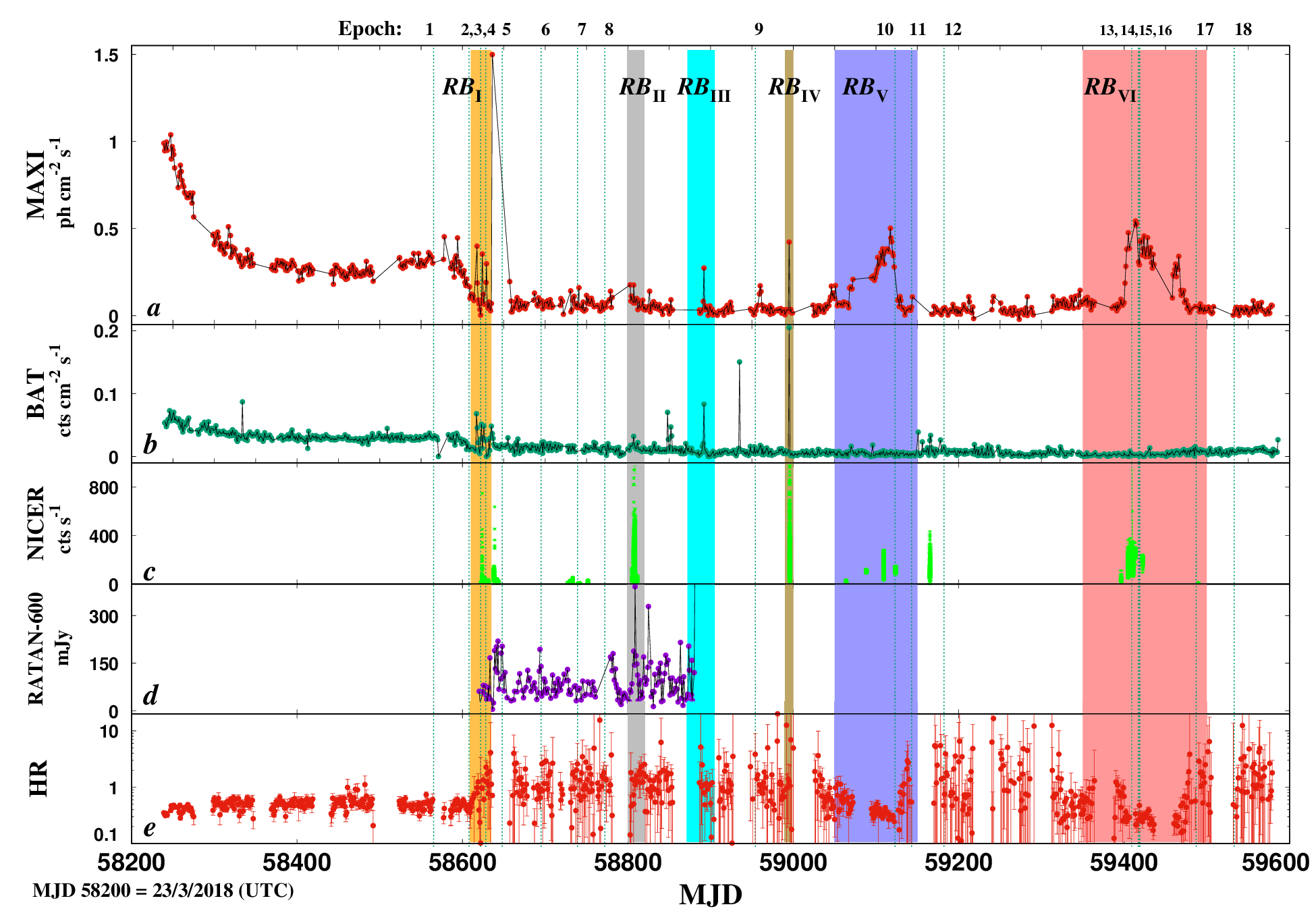}

\caption{The \textit{MAXI} flux (2 -- 20 keV) of the source between the period  MJD 58200 to MJD 59600 is shown in the top-panel (panel \textit{a}). The panels \textit{b, c} and \textit{d} show the \textit{BAT} flux (15 - 50 keV), \textit{NICER} flux (0.5 -- 12 keV) and the  {radio flux from RATAN-600 at 11.2 GHz band (mJy)}, respectively. Panel \textit{e} shows the HR (6 -- 20 keV / 2 -- 6 keV) obtained using the \textit{MAXI} data. The shaded regions in gold, grey, cyan, olive-green, blue and pink colours represent the re-brightening phases - RB$_{\rm I}$, RB$_{\rm II}$, RB$_{\rm III}$, RB$_{\rm IV}$, RB$_{\rm  V}$ and RB$_{\rm VI}$ respectively. The numbered vertical green dotted lines mark the 18 simultaneous broadband observations made by \textit{AstroSat}, \textit{NICER} and \textit{NuSTAR}.  {The \textit{MAXI}, \textit{BAT} and \textit{NICER} light curves have a bin time of 1 day, 1 day and 1 sec, respectively.} \label{fig 2A}}
\end{figure*}

\subsection{Spectral Analysis and Modeling}
\label{sec 3.2}

The spectral analyses for all the observations were done using \texttt{XSpec v12.10.1f} and and \texttt{HEASOFT v6.29}. We perform the broadband spectral analysis and modeling across 0.7 -- 60 keV using simultaneous data from \textit{SXT} (0.7 -- 7 keV) + \textit{LAXPC} (3 -- 60 keV) and \textit{NICER} (0.7 -- 12 keV) + \textit{NuSTAR} (3 -- 60 keV). The difference in flux normalization between two instruments, while performing the simultaneous fit, is taken care by the multiplicative model, \textit{constant}. Absorption due to the interstellar medium (ISM) of the Galaxy has been modeled using the \textit{TBabs} model with all the abundance parameters set to default as mentioned in \citealt{2000ApJ...542..914W}. 
The combined spectra were initially fitted using the multi-temperature disc blackbody (\textit{diskbb}) \citep{1984PASJ...36..741M, 1986ApJ...308..635M} model and the Comptonisation (\textit{nthComp}) \citep{1996MNRAS.283..193Z, 1999MNRAS.309..561Z} model individually. The spectra corresponding to Epochs 6 -- 9, 11, 12, 17 and 18 showed acceptable fits using the \textit{nthComp} model only, whereas Epochs 1 -- 5, 10 and 13 -- 16 required the combination of both models, \textit{diskbb+nthComp} for the continuum.  {We also use a partial covering absorption model, \textit{TBpcf} along the continuum for all the observations, considering the recent evolution of the source into the `obscured' phase (see \citealt{2021MNRAS.503..152M, 2022MNRAS.510.3019A}). \textit{TBpcf} addresses the edge at $\sim7$ keV which is interpreted as a neutral iron K-alpha photoelectric edge. This spectral feature, conventionally seen in AGNs, is described by the partial covering absorption by winds/gas clouds \citep{2002MNRAS.329L...1B}.} \textit{TBpcf} quantifies the equivalent hydrogen column density that is local to the source (N$_{\rm H_{\rm 1}}$) and the covering fraction (PCF) of the obscuration (see also \citealt{2017MNRAS.471.1797M}). Henceforth, the model combinations -  \textit{TBabs(TBpcf(diskbb+nthComp))} and \textit{TBabs(TBpcf(nthComp))} will be referred as Model-1 and Model-2 respectively. Along with the above mentioned model combinations, few additional models were required to address the absorption and emission features between 6 -- 9 keV. For example, the \textit{gaussian} model was used to address the emission line between $6 - 8$ keV. Epochs 13 and 17 showed a narrow absorption feature in the same energy range, which was addressed by the \textit{gabs} \citep{2010ApJ...710.1755N} model. A broad absorption feature was also present in few Epochs (Epochs 6, 7, 8, 11, 12 and 13) between 7 -- 9 keV, for which the \textit{smedge} \citep{1994PASJ...46..375E} model was used. The additional Si and Au edges in the \textit{SXT} spectra were addressed using the \texttt{gain fit} command (see \citealt{2022MNRAS.510.3019A} and the references therein), with the slope set to 1. We use the \textit{edge} model to address the instrumental Xenon edge ($\sim32$ keV) observed in the \textit{LAXPC} spectrum \citep{2017ApJS..231...10A, 2019MNRAS.487..928S}.

 {The \textit{NICER} spectra (0.7 -- 12 keV) corresponding to the re-brightening phases (Table \ref{tab 2}) were initially fitted  using Model-1, considering the requirement of disc and  Comptonization model components to produce a best fit for the broadband observations. Nonetheless, all the \textit{NICER} observations required only a single component for the continuum, except  Obs. 1 corresponding to RBI, where this particular observation showed improved fit with a combination of 
\textit{diskbb} and \textit{nthComp} model components (Model-1). However, for this observation, we had to freeze both disc parameters to the values close to the ones obtained from the broadband best-fits. The remaining \textit{NICER} observations did not show the requirement of two components for the continuum.  We then tried to fit the remaining \textit{NICER} spectra using the \textit{diskbb} model component along with \textit{TBabs} and \textit{TBpcf}. While few of the observations produced good fits, the rest of the observations produced non-physical disc temperatures. Following \cite{2020ApJ...902..152N}, we also attempted to fit the \textit{NICER} spectra using either with the \textit{powerlaw} model or  with the \textit{bbody} model. Although these models produced satisfactory fits for few observations, they turned non-viable for most of the observations. Additionally, we also tried \textit{simpl*diskbb} to fit the observations. Unfortunately, that model combination also did not work for most of the observations. And for the ones it did, we got an extremely steep photon index ($\Gamma~>$ 4), with high errors. At face value, it can be broadly inferred that the \textit{NICER} spectra of the source in the 0.7 – 12 keV energy band, corresponding to the obscured phase, is best described using a single Comptonized component and does not show the requirement for an additional disc component. We infer that this could be because of the fact that the disc temperature in the \textit{nthComp} model component adequately describes the moderately faint disc flux, without actually having to include the disc component. We, therefore, proceeded to fit all the \textit{NICER} observations using Model-2, which produced acceptable fits for all the observations.} All the \textit{NICER} spectra also showed additional absorption and emission features, which are addressed using the \textit{gaussian, gabs} and \textit{smedge} models. Errors for the parameters are estimated using the error command in \texttt{XSpec}. All the error values are quoted at 90\% confidence level.  {However, the error values of certain parameters were too small ($< 5\%$), and thus are tagged with a dagger symbol in the tables (Tables \ref{tab 3}, \ref{tab 5}, \ref{tab 4} and \ref{tab 6}) to notify that the error values are insignificant.}

We computed the unabsorbed bolometric luminosity (L$_{\rm bol}$) in 0.3 -- 100 keV energy range using the relation, L$_{\rm bol}$ = 4$\pi$D${^2}\times (F)$, where D is the distance of the source (D = 8.2 kpc \citep{2014ApJ...796....2R}; also see \citealt{2022MNRAS.510.3019A} and the references therein) and the \textit{F} is the unabsorbed intrinsic flux (0.3 -- 100 keV), which is estimated by incorporating the \textit{cflux} model along the continuum. The partial covering absorption model (\textit{TBpcf}) is excluded while integrating the \textit{cflux} model along with the continuum components. The flux value thus obtained does not account the effects of obscuration. For example, we used the resultant model combination \textit{TBabs(TBpcf(cflux(diskbb+gauss+nthComp)))} to estimate the unabsorbed intrinsic flux during Epoch 10. 

In the subsequent section, we present the results obtained from our analysis.

\section{Results}
\label{sec 4}

\subsection{Prolonged Low Luminosity Phase}
\label{sec 4.1}

In Figure \ref{fig 2A}, we show the flux variation of GRS 1915+105, starting from  March 2019 to  {November 2021}, as observed by multiple instruments both in the X-rays and the radio bands. The top four panels display the source flux as observed by \textit{MAXI} (2 -- 20 keV), \textit{BAT} (15 -- 50 keV), \textit{NICER} (0.3 -- 10 keV) and \textit{RATAN-600} radio telescope (11.2 GHz) respectively, while the bottom-most panel shows the Hardness Ratio (HR = (6 -- 20 keV) / (2 -- 6 keV)) of the source obtained from \textit{MAXI} light curve. 

GRS 1915+105 is observed in the decay phase at the beginning of the observation period, where \textit{MAXI} flux showed a decrease from $\sim$1 ph cm$^{-2}$ s$^{-1}$ during MJD 58250 to $\sim0.4$ ph cm$^{-2}$ s$^{-1}$ during MJD 58400. This pattern is also reflected in the \textit{BAT} light curve, where the flux does not drastically decrease, rather a marginal decrease is seen from 0.3 cts cm$^{-2}$ s$^{-1}$ to 0.15 cts cm$^{-2}$ s$^{-1}$. There is also a gradual increase in the HR from 0.4 -- 0.7 during this period. The source exhibited a consistently low flux since MJD 58600. However, this low luminosity phase is noticed to be intermittently perturbed by strong and suḍden re-brightenings in X-rays. We refer to these sudden re-brightenings as the re-brightening (RB) phases. The six re-brightening phases, referred as RB$_{\rm I}$, RB$_{\rm II}$, RB$_{\rm III}$, RB$_{\rm IV}$, RB$_{\rm V}$, and RB$_{\rm VI}$ are shown in gold, grey, cyan, olive-green, blue, and pink colour-shaded regions respectively in Figure \ref{fig 2A}.

The \textit{MAXI} light curve of GRS 1915+105 showed a sequence of oscillations with flux varying between 0.02 to 0.4 ph cm$^{-2}$ s$^{-1}$ during re-brightening phase I (RB$_{\rm I}$) between MJD 58600 to 58633. 
 {RB$_{\rm II}$ (MJD 58799) and RB$_{\rm IV}$ (MJD 58994) represent quick flares lasting for $\sim900$ and 2500 sec, respectively. RB$_{\rm III}$ (MJD 58891) is also recognized as a quick flare. But due to the lack of observations of the source during RB$_{\rm III}$, the exact duration of the quick flare could not be determined.} The source exhibited a sudden increase in the \textit{MAXI} flux from $\sim$0.02 to 0.5 ph cm$^{-2}$ s$^{-1}$ during these quick flares. The available radio data also reveals that RB$_{\rm I}$, RB$_{\rm II}$ and RB$_{\rm III}$ were precursory to the radio flares (see panel \textit{d} of Figure \ref{fig 2A}). In addition to the quick flares, GRS 1915+105 also exhibited two relatively prolonged re-brightenings  (RB$_{\rm V}$ and RB$_{\rm VI}$), which lasted for $\sim$100 days and $\sim$150 days respectively. Source showed a gradual increase and decrease in the flux analogous to a mini-outburst \citep{2012MNRAS.423.3308J, 2020MNRAS.496.1001C} and the average \textit{NICER} flux varied from $\sim$20 cts s$^{-1}$ at the beginning to $\sim120$ cts s$^{-1}$ at the peak of RB$_{\rm V}$ and $\sim$30 cts s$^{-1}$ to $\sim250$ cts s$^{-1}$ from the beginning to the peak of RB$_{\rm VI}$ (see Table \ref{tab 2}). The drop in HR during both RB$_{\rm V}$ and RB$_{\rm VI}$ indicates a slow propagation of the source towards the softer state during these two re-brightenings. All these re-brightening phases are extensively studied using the \textit{NICER} observations. RB$_{\rm I}$, RB$_{\rm V}$ and RB$_{\rm VI}$ have also been observed in the broad energy band by \textit{AstroSat} and \textit{NuSTAR}. Although,  {RB$_{\rm II}$} and  {RB$_{\rm IV}$} being quick and sudden flares, could not be monitored in the broadband energies. RB$_{\rm III}$ is not studied in this paper due to the lack of observations.

\subsection{The Re-brightening Phases}
\label{sec 4.2}

\subsubsection{Re-brightening phase I ({RB$_{\rm I}$})}
\label{sec 4.2.1}
 
\begin{figure*}
\centering
\hspace*{-0.6cm} \includegraphics[width=9.2cm, height=8cm]{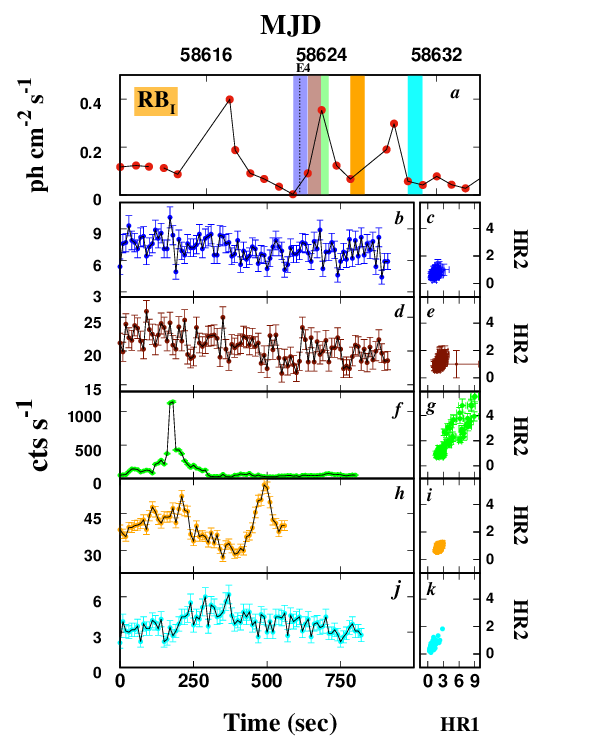} 
\hspace*{-0.8cm}\includegraphics[width=9cm, height=8.2cm]{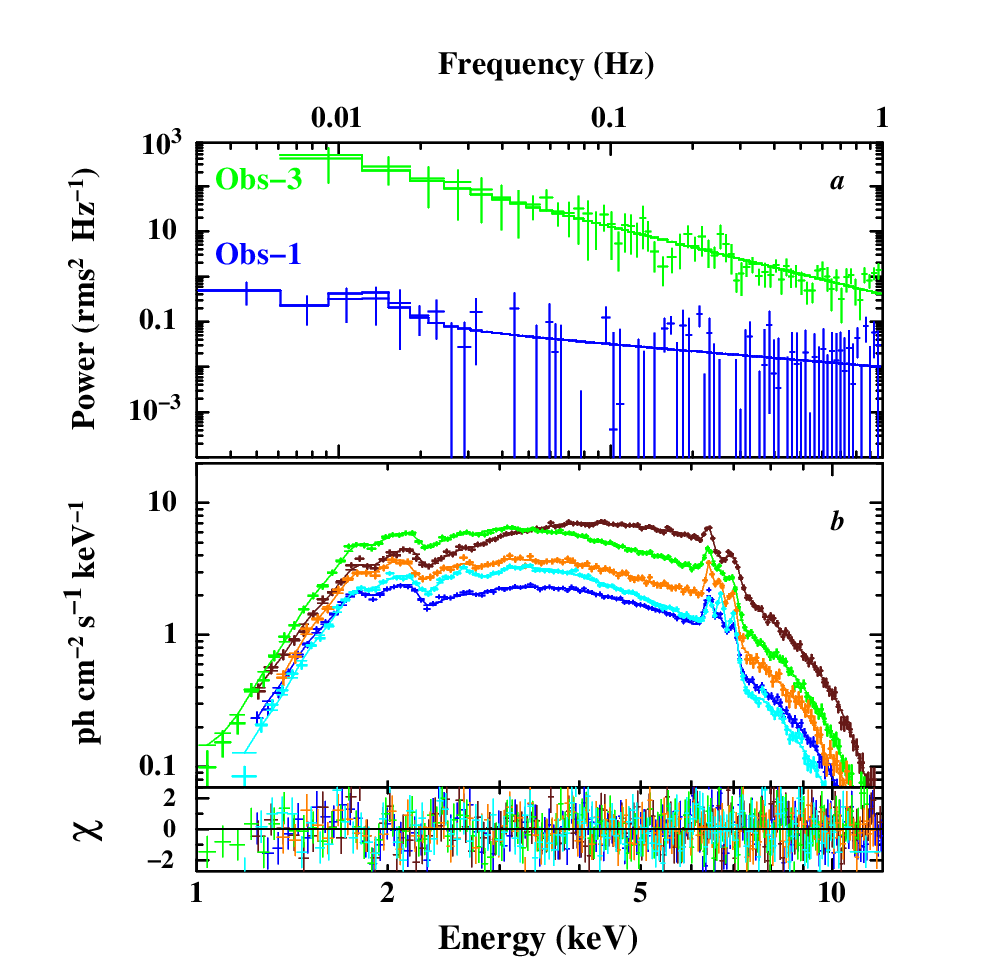}
\caption{\textit{Left (Figure \ref{fig 3}a)}: Panel \textit{a} shows the \textit{MAXI} light curve of the source during the first phase of re-brightening - RB${_{\rm I}}$ (7 May 2019 to 6 June 2019).  {The black dashed line in this panel corresponds to the simultaneous broadband\textit{ NICER+NuSTAR} observation (Epoch 4).} The 5 \textit{NICER} observations corresponding to the low phase (Obs. 1), rise (Obs. 2), peak (Obs. 3), decay (Obs. 4) and the low phase after the decay (Obs. 5) are shown in blue, brown, green, orange and cyan color-shaded regions respectively.  {Panels  \textit{b \& c, d \& e, f \& g, h \& i} and \textit{j \& k} show the \textit{NICER} light curves \& CCDs corresponding to Obs. 1, 2, 3, 4 and 5 in the above-mentioned colors.} \textit{Right (Figure \ref{fig 3}b)}:  Panel \textit{a} shows the overplot of PDS corresponding to Obs. 1 and 3, plotted in blue and green colors respectively. The PDS plotted during Obs. 3 is scaled by a factor of 10$^{4}$ w.r.t. that of Obs. 1.  {Panel \textit{b} shows the overplot of the energy spectra along with the residuals from the corresponding spectral fits obtained from all the 5 \textit{NICER} observations.} See text for details. 
\label{fig 3}}
\end{figure*}

The \textit{MAXI} light curve in the top-panel of Figure \ref{fig 3}a shows the flux variation of the source during RB$_{\rm I}$, starting from MJD 58600 to MJD 58633. The 5 color-shaded regions in the top-panel represent the 5 \textit{NICER} observations that  we considered to study RB$_{\rm I}$, where each observation corresponds to a different phase of the re-brightening. Obs. 1 corresponds to the low luminosity phase (blue color-shaded region in the top-panel of Figure \ref{fig 3}a). Obs. 2, 3, 4 and 5 correspond to the rise (shaded in brown), flare (shaded in green), decay (shaded in orange) and low phase after the decay (shaded in cyan) respectively.  {The \textit{NICER} light curves corresponding to all the 5 observations are shown in panels \textit{b, d, f, h \& j} respectively, while the corresponding CCDs are shown in the adjacent panels (\textit{c, e, g, i \& k} respectively).}
The above-mentioned colour index is followed throughout this paper, while plotting the \textit{NICER} data points corresponding to each phase. The \textit{NICER} flux varied from 7 cts  s$^{-1}$ at the low phase to 1200 cts  s$^{-1}$ at the peak of the re-brightening. The variation in the average count rate during each observation is shown in Table \ref{tab 2}. In the CCD, during Obs. 1, 2, 4 and 5, HR1 and HR2 varied between the limits 0.1 $\leq$ HR1 $\leq$ 5 and 0.2 $\leq$ HR2 $\leq$ 2. However, during Obs. 3 (flare), the  {upper value of the range} of both HR1 and HR2 increased significantly to 9 and 6 respectively. These model independent analysis does not explicitly unveil any variability classes in the source during  RB$_{\rm I}$.

In the top-panel of Figure \ref{fig 3}b, we show an overplot of the PDS obtained from Obs. 1 and 3, plotted in blue and green points respectively. The PDS during Obs. 2, 4 and 5 is dominated with broadband noise beyond 0.05 Hz, similar to the PDS corresponding to Obs. 1 shown in the figure. PDS obtained from Obs. 3 showed a power-law distribution. The total rms varied between 9$_{-2}^{+1}$\% to 44$_{-4}^{+3}$\% in 0.003 -- 1 Hz frequency range.  {The bottom-panel of the Figure \ref{fig 3}b shows an overplot of spectra obtained from all the 5 observations, along with the residuals from the best-fit model. The spectral analysis for all the 5 observations is carried out with Model-2. The photon index ($\Gamma$) varies between $1.29_{-0.04}^{+0.02}$ -- $1.47_{-0.02}^{+0.04}$ throughout RB$_{\rm I}$. The N$_{\rm H}$ value varied between $5.0_{-0.2}^{+0.1} - 5.5_{-0.3}^{+0.2} \times 10^{22}$ atoms cm$^{-2}$. Along with interstellar absorption (N$_{\rm H}$) all the 5 observations also showed effects of local obscuration. We observed an additional column density (N$_{\rm H_{\rm 1}}$) of $82_{-4}^{+4}\times10^{22}$ atoms cm$^{-2}$ initially, which decreased to $18^{-1}_{+2}\times10^{22}$ during the flare. N$_{\rm H_{\rm 1}}$ further increased to $108^{-5}_{+8}\times10^{22}$ at the end of RB$_{\rm I}$. The PCF however, varied randomly between 0.56 -- 0.77. The best-fitted timing and spectral parameters are mentioned in Tables \ref{tab 3} and \ref{tab 4} respectively.}

\subsubsection{Re-brightening phase II and IV  {(RB$_{\rm II}$ \& RB$_{\rm IV}$)}}
\label{sec 4.2.2}

RB$_{\rm II}$ and RB$_{\rm IV}$ (grey and olive-green shaded regions in Figure \ref{fig 2A}) are two quick flares that spanned for $\sim$1000 sec and 2700 sec respectively. The \textit{MAXI} flux corresponding to RB$_{\rm II}$ is partly missing, while the \textit{NICER} flux showed a steady flux with an average value of 38 cts sec$^{-1}$. However, the radio observations (panel \textit{d} of Figure \ref{fig 2A}) show a simultaneous radio flare with the flux varying from $\sim40$ to 400 mJy. During RB$_{\rm IV}$, the \textit{MAXI} flux is suddenly seen increasing from 0.04 to 0.5 ph cm$^{-2}$ s$^{-1}$. This flare pattern is also observed in the \textit{BAT} and the \textit{NICER} light curves. The hardness during both RB$_{\rm II}$ and RB$_{\rm IV}$ is relatively higher with the HR values > 1.5 (panel \textit{e} of Figure \ref{fig 2A}).

\begin{figure*}
\centering
\hspace*{-0.6cm} \includegraphics[width=9.2cm, height=8cm]{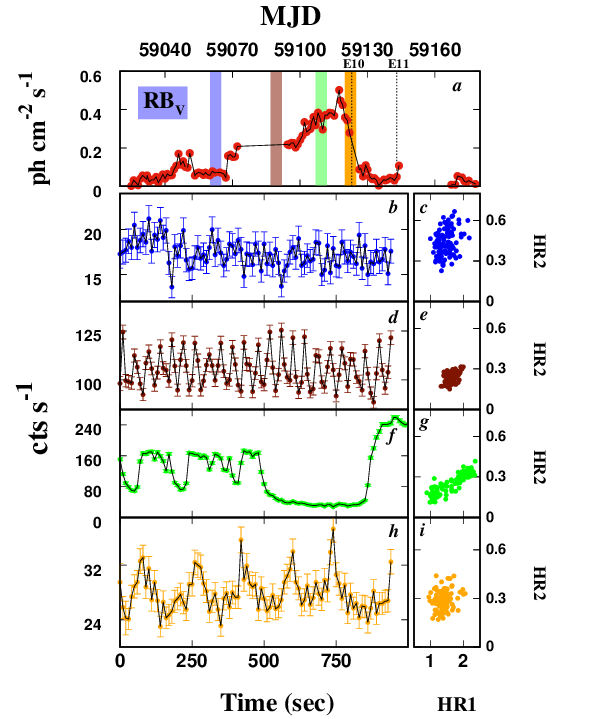}
\hspace*{-0.8cm} \includegraphics[width=9cm, height=8.2cm]{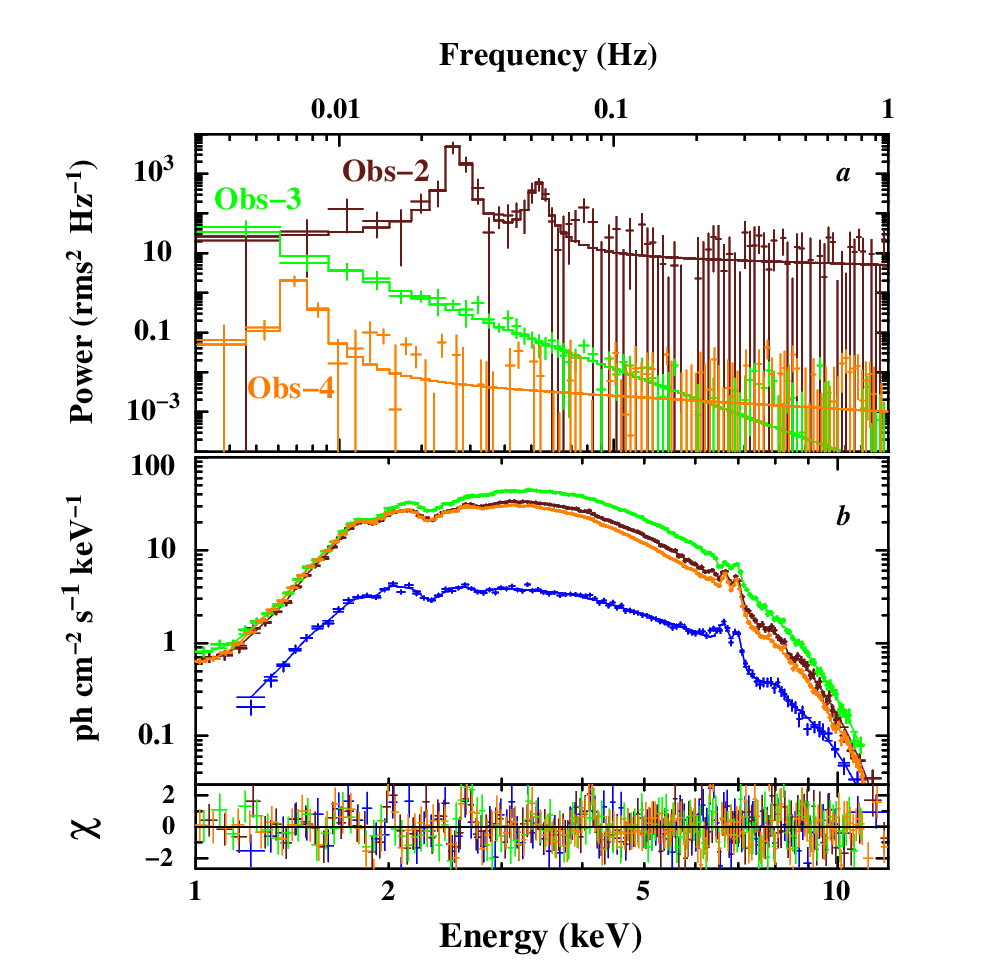}
\caption{\label{fig 4} \textit{Left (Figure \ref{fig 4}a)}: The \textit{MAXI} light curve of the source during RB${_{\rm V}}$ (20 July 2020 -- 28 October 2020) is shown in panel \textit{a}.  {The two black dashed lines correspond to the simultaneous wide-band observations from Table \ref{tab 1}, E10 and E11.} The 4 \textit{NICER} observations corresponding to the low (Obs. 1), rise  (Obs. 2), decay  (Obs. 3) and the low phase after the decay  (Obs. 4) of RB${_{\rm V}}$, are shaded in blue, brown, green, and orange respectively.  {Panels \textit{b \& c, d \& e, f \& g} and \textit{h \& i} show the \textit{NICER} light curves \& CCDs corresponding to Obs. 1, 2, 3 and 4.} \textit{Right (Figure \ref{fig 4}b)}: Panel \textit{a} shows the overplot of PDS obtained from Obs. 2, 3 and 4, plotted in brown, green and orange respectively. PDS obtained from Obs. 2 shows QPO and $1^{\rm st}$ Harmonic at 26 and 52 mHz, while PDS obtained from Obs. 4 shows a QPO at 6 mHz. The PDS corresponding to Obs. 2 and 3 are scaled by a factor of 10$^{3}$ and 10$^{6}$ respectively, w.r.t. that of Obs. 4.  {Panel \textit{b} shows the overplot of the energy spectra  along with the residuals from the corresponding spectral fits of all the 4 observations.}  See text for details.}
\end{figure*}

 {The power spectra obtained from RB$_{\rm II}$ and RB$_{\rm IV}$ is characterized by a power law distribution. No QPO features were identified and the PDS was dominated with broadband noise above 0.1 Hz for both re-brightening phases. Model-2 provides the best-fit for spectra from both RB$_{\rm II}$ and RB$_{\rm IV}$. The source showed a $\Gamma$ value of 1.20$_{-0.03}^{+0.02}$ and 1.16$_{-0.06}^{+0.06}$ for RB$_{\rm II}$ and RB$_{\rm IV}$ respectively. A constant galactic absorption with N$_{\rm H} ~ \sim ~ 5.5\times10^{22}$ atoms cm$^{-2}$ was observed during both observations. The additional column density drastically varied between RB$_{\rm II}$ and RB$_{\rm IV}$, with N$_{\rm H_{\rm 1}}\sim16^{+3}_{-2}\times 10^{22}$ atoms cm$^{-2}$ and a PCF of 0.63$_{-0.02}^{+0.03}$ for RB$_{\rm II}$ and N$_{\rm H_{\rm 1}}\sim78^{+4}_{-7}\times 10^{22}$ atoms cm$^{-2}$ and a PCF of 0.71$_{-0.02}^{+0.02}$ for RB$_{\rm IV}$. The results of the fit are presented in Tables \ref{tab 3} and \ref{tab 4}.}

\subsubsection{Re-brightening phase V ({RB$_{\rm V}$})}
\label{sec 4.2.3}

The prolonged re-brightening phase, RB$_{\rm V}$, spanned from MJD 59050 to 59150 with the  \textit{MAXI} flux varying from $\sim0.02$ ph cm$^{-2}$ s$^{-1}$ at the beginning to $\sim0.6$ ph cm$^{-2}$ s$^{-1}$ at the peak of RB$_{\rm V}$ (top-panel of Figure  \ref{fig 4}a).  {The \textit{NICER} light curves and the corresponding CCDs pertaining to the low (Obs. 1), rise (Obs. 2), peak (Obs. 3) and the decay (Obs. 4) phases of RB$_{\rm V}$ are shown in the bottom panels (\textit{b \& c, d \& e, f \& g} and \textit{h \& i}, respectively).} At the beginning of RB$_{\rm V}$, the source displayed low flux ($\sim$ 15 cts sec$^{-1}$) with no structured variability in the light curve. Recurring burst profiles with a periodicity of $\sim$ 50 sec, were observed in the light curve during the rise. Each flare profile showed a varying peak amplitude. These flares also did not resemble the typical heart-beat profile ($\rho$ class)  {as classified by \citealt{2000A&A...355..271B}}. We therefore, classify the source belongs to $\rho^\prime$ variability class (defined as a variant of the $\rho$ class; see \citealt{2022MNRAS.510.3019A}), during Obs. 2. At the peak of RB$_{\rm V}$, the source exhibited large amplitude variability with flux varying between 40 cts sec$^{-1}$ at the dip to 220 cts sec$^{-1}$ at the peak of the variability. The source variability during Obs. 3 resembled with the $\lambda$ variability class (\citealt{2000A&A...355..271B}).  {During the decay, the source is seen exhibiting the typical $\rho$ profile with a periodicity of $\sim$160 sec. Obs. 1 exhibits high HR values in the CCD with HR2 going as high as 0.75 (panel \textit{c} in Figure \ref{fig 4}a), while Obs. 2, 3 and 4 show relatively lower HR values, HR2 < 0.4 (see panels \textit{e}, \textit{g} and \textit{i} in Figure \ref{fig 4}a).} The relatively higher HR values and the absence of any variability structure in the light curve during Obs. 1 leads to the assumption that the source exhibits $\chi$ variability class during Obs. 1. 

The top-panel of Figure \ref{fig 4}b shows an overplot of the PDS obtained from Obs. 2, 3 and 4 plotted in brown, green and orange colors respectively. The PDSs show QPOs at 26 mHz and 6 mHz during Obs. 2 and 4,  attributing to the periodicity of the $\rho^\prime$ and $\rho$ profiles in the light curves. Obs. 3 showed a powerlaw noise distribution in the PDS, while the PDS during Obs. 1 was dominated with noise beyond 0.1 Hz. The total rms of the source varied between 8.2$_{-0.8}^{+0.9}$\% -- 40.2$_{-2.1}^{+4.9}$\%.

 {The \textit{NICER} spectra corresponding to all 4 observations are well described with Model-2. $\Gamma$ increases as the source proceeds from the low phase (Obs. 1) to the peak phase (Obs. 3) from 1.34$_{-0.04}^{+0.03}$ to 2.47$_{-0.06}^{+0.03}$ and decreases down to 1.91$_{-0.01}^{+0.03}$ during Obs. 4. The source showed an almost constant galactic hydrogen column density with N$_{\rm H}$  $\sim5.0\times10^{22}$ atoms cm$^{-2}$. A high N$_{\rm H_{\rm 1}}$ of 150$^{+12}_{-10}\times10^{22}$ atoms cm$^{-2}$ is observed during the low phase (Obs. 1), while N$_{\rm H_{\rm 1}}$ drastically drops to $\sim4^{-0.4}_{+0.3}-8^{-0.1}_{+0.1}\times10^{22}$ atoms cm$^{-2}$ during Obs. 2, 3 and 4. The PCF varied between 0.60 to 0.79$_{-0.03}^{+0.03}$ without a pattern. In the bottom-panel of Figure \ref{fig 4}b, we present an overplot of the fitted \textit{NICER} spectra for all 4 observations along with the residuals obtained after fitting with Model-2. All the model fitted parameters are summarized in Tables \ref{tab 3} and \ref{tab 4}. }

\subsubsection{Re-brightening phase VI ({RB$_{\rm VI}$})}
\label{sec 4.2.4}

\begin{figure*}
\centering
\hspace*{-0.6cm} \includegraphics[width=9.2cm, height=8.2cm]{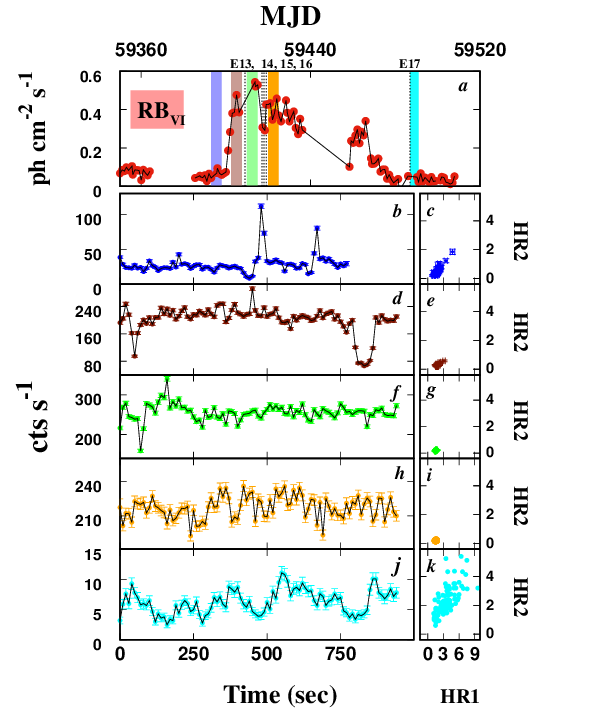}
\hspace*{-0.8cm} \includegraphics[width=9.2cm, height=8.5cm]{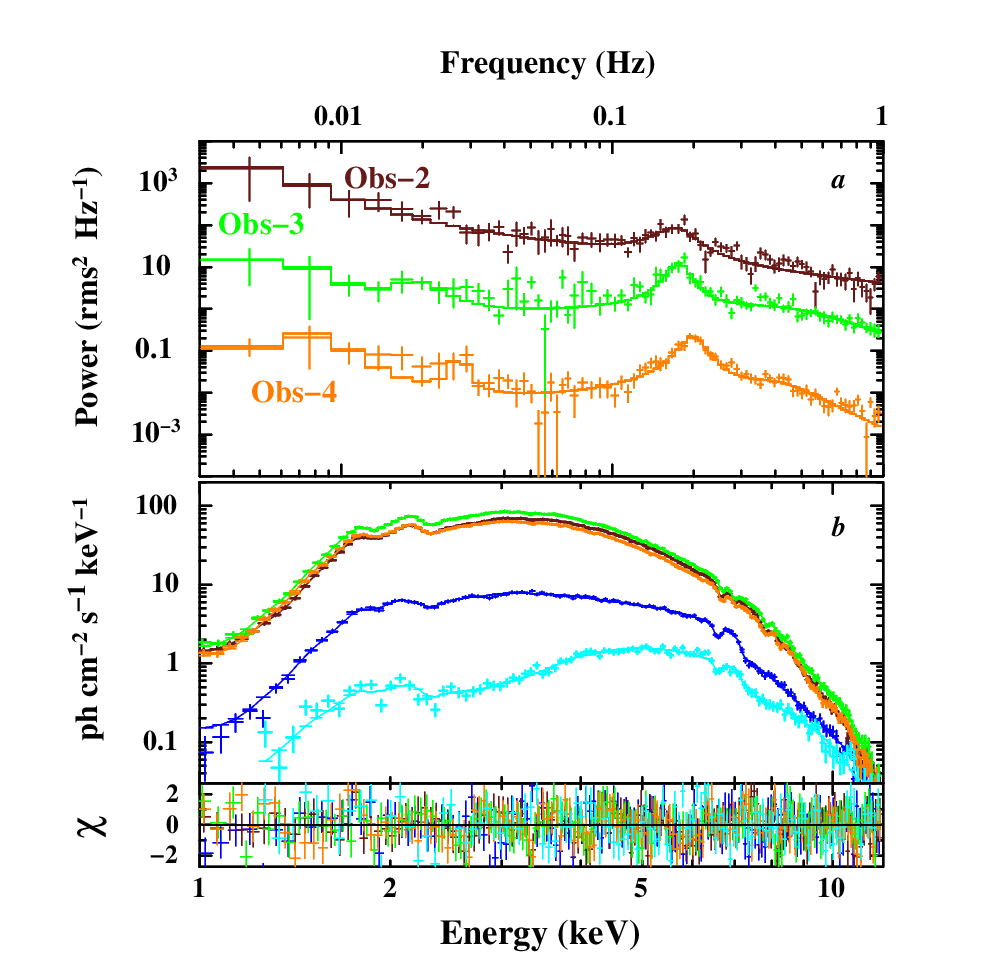}
\caption{\textit{Left (Figure \ref{fig 5}a)}: The \textit{MAXI} flux of GRS~1915+105 during RB${_{\rm VI}}$ (15 June 2021 -- 13 October 2021) is shown in panel \textit{a}.  {The simultaneous wide-band observations, Epochs $13 - 17$ from Table \ref{tab 1}, is marked by black dashed lines in panel \textit{a}.} {Panels - \textit{b \& c, d \& e, f \& g, h \& i} and \textit{j \& k} show the \textit{NICER} light curves \& CCDs corresponding to the 5 phases of the re-brightening: low, rise, peak, decay and the low phase after the decay respectively.} \textit{Right (Figure \ref{fig 5}b)}: An overplot of PDS obtained from Obs. 2, 3 and 4 (plotted in brown, green and orange respectively) is shown in panel \textit{a}. All the three PDS shows QPO features at 0.17, 0.18 and 0.2 Hz respectively. The PDS corresponding to Obs. 2 and 3 are scaled by a factor of 10$^{4}$ and 10$^{2}$ respectively, w.r.t that of Obs. 4.  {An overplot of the energy spectra of all the 5 observations, along with the residuals from the corresponding best-fit models is shown in panel \textit{b}}. See text for details. \label{fig 5}}
\end{figure*}

RB$_{\rm VI}$, observed from MJD 59350 to 59500, is also a prolonged re-brightening phase like RB$_{\rm V}$. However, RB$_{\rm VI}$ portrayed a fast-rise and slow decay light curve profile, in contrast to RB$_{\rm V}$ that showed a slow-rise and fast-decay profile in the light curve. The flux evolution of the source during RB$_{\rm VI}$ is shown in the \textit{MAXI} light curve (top-panel of Figure \ref{fig 5}a). The light curves obtained from all the 5 observations did not show any periodic variability structure  {(see panels \textit{b, d, f, h} and \textit{j} in Figure \ref{fig 5}a)}.  {The CCDs corresponding to Obs. 1 -- 4 showed moderate HR values with HR1 < 3 and HR2 < 0.8 (panels \textit{c, e, g} and \textit{i} in Figure \ref{fig 5}a). But Obs. 5 shows an increased hardness with  {the upper value of the range} of HR1 and HR2 extending up to 9 and 6 respectively (panel \textit{k} in Figure \ref{fig 5}a).} The avg. count rate corresponding to each observation is given in Table \ref{tab 2}.

The PDS obtained from Obs. 2, 3, and 4 (top-panel of Figure \ref{fig 5}b) showed QPOs at 170, 180 and 200 mHz respectively, with  {Q-factor} evolving from 2.70$_{-0.01}^{+0.02}$ (Obs. 2) to 4.22$_{-0.03}^{+0.03}$ (Obs. 4). The total rms varied between 15$_{-1}^{+1}$\% -- 20$_{-2}^{+3}$\% during Obs. 2, 3 and 4. PDS corresponding to Obs. 1 and 5 exhibited high fractional variability (> 26\%) and showed no indication of QPOs. The parameters are provided in Table \ref{tab 3}.

 {We present an overplot of the modeled spectra  corresponding to the 5 \textit{NICER} observations along with the residuals in the bottom-panel of Figure \ref{fig 5}b. All the 5 spectra were well-fitted using Model-2. Initially, source showed $\Gamma$ of 1.36$_{-0.02}^{+0.02}$ during Obs. 1. The spectra showed a steeper $\Gamma$ value of 2.04$_{-0.04}^{+0.02}$ during Obs. 3, which again decreased to 1.37$_{-0.4}^{+0.04}$ during Obs. 5. A constant kT$_{\rm e}$of $\sim 1.9$ keV was observed throughout RB$_{\rm VI}$. N$_{\rm H_{\rm 1}}$ value was minimum during the rise, peak and the decay phases, with values varying between $4.4_{-0.4}^{+0.6}-8.9_{-0.6}^{+0.6}\times10^{22}$ atoms cm$^{-2}$. An  increased N$_{\rm H_{\rm 1}}$ was observed during Obs. 1 (N$_{\rm H_{\rm 1}}\sim98_{-10}^{+5}\times10^{22}$ atoms cm$^{-2}$) and Obs. 5 (N$_{\rm H_{\rm 1}}\sim44_{-4}^{+6}\times10^{22}$ atoms cm$^{-2}$).  {With reference to  \citealt{2000A&A...355..271B},} and based on the CCD, PDS and the spectral characteristics, the source could possibly belong to the hard state (or the $\chi$ class) during the beginning and the end of RB$_{\rm VI}$, while it exhibited the $\delta$ variability classes during the rise, peak and the decay phases. The best-fitted model parameters are given in Tables \ref{tab 3} and \ref{tab 4}.}

\begin{table*}
\centering
\caption{Details of model fitted parameters of power spectra obtained from {\it NICER} observations of GRS~1915$+$105 in $0.5 - 12$ keV energy band for each segment during each Re-brightening phases (see Table \ref{tab 2}). Multiple \textit{Lorentzians} were used to fit the PDS.  Here, $\nu_{\rm QPO}$ represents the QPO frequency in Hz, $\sigma$ denotes the QPO significance and the total rms (rms$_{\rm Tot}$), given in percentage, is estimated for 0.003 -- 1 Hz frequency range. Errors for all the parameters are calculated with 90\% confidence.}

\begin{tabular}{|lcccccc|}
\hline
Epoch & $\nu_{\rm QPO}$ & Q-factor & $\sigma$ & rms$_{\rm QPO}$ &  rms$_{\rm Tot}$ &  {$\chi^{2}/d.o.f$}\\
& (Hz) & & & (\%) &  (\%) & \\
\hline
\multicolumn{7}{|c|}{RB$_{\rm I}$} \\
\hline
\rule{0pt}{10pt}Obs. 1 & - & - & - & -  & 15.24$_{-1.04}^{+0.92}$ &  {95.19/80}\\
Obs. 2 & - & - & - & - & 9.24$_{-2.41}^{+0.94}$ &  {107.48/82} \\
Obs. 3 & - & - & - & - & 43.82$_{-3.67}^{+3.01}$ &  {162.81/154} \\ 
Obs. 4 & - & - & - & - & 13.78$_{-0.88}^{+0.84}$ &  {112.90/76}\\ 
Obs. 5 & - & - & - & - & 20.92$_{-1.02}^{+1.00}$ &  {108.3/99}\\
\hline
\multicolumn{7}{|c|}{RB$_{\rm II}$} \\
\hline
Obs. 1 & - & - & - & - & 22.87$_{-1.46}^{+1.81}$ &  {122.6/113}\\ 
\hline
\multicolumn{7}{|c|}{RB$_{\rm IV}$} \\
\hline
Obs. 1 & - & - & - & - & 46.42$_{-4.06}^{+3.11}$ &  {104.2/119}\\ 
\hline
\multicolumn{7}{|c|}{RB$_{\rm V}$} \\
\hline
Obs. 1 & - & - & - & - & 13.14$_{-0.79}^{+1.10}$ &  {102.9/102}\\
Obs. 2 & 0.02$\dagger$ & 16.30$_{-1.28}^{+1.65}$  & 2.76  & 7.77$_{-0.94}^{+0.81}$  & 8.23$_{-0.81}^{+0.89}$ &  {81.86/76}\\
Obs. 3 & - & - & - & - & 40.23$_{-2.12}^{+4.94}$ &  {94.86/80}\\
Obs. 4 & 0.01$\dagger$ & 12.12$_{-0.91}^{+0.92}$ & 2.80  & 6.77$_{-0.46}^{+0.68}$  & 13.78$_{-1.51}^{+1.06}$ &  {109.84/97} \\  
\hline
\multicolumn{7}{|c|}{RB$_{\rm VI}$} \\
\hline 
Obs. 1 & - & - & - & - & 32.02$_{-3.22}^{+2.90}$ &  {108/79} \\
Obs. 2 & 0.17$\dagger$ & 2.70$_{-0.01}^{+0.02}$ & 7.44 & 9.54$_{-0.41}^{+0.39}$ & 20.04$_{-2.51}^{+2.73}$ &  {91.13/75} \\
Obs. 3 & 0.18$\dagger$ & 5.33$_{-0.04}^{+0.04}$ & 6.31 & 12.26$_{-0.98}^{+1.42}$ & 17.18$_{-1.40}^{+1.29}$  &  {99.4/70}\\
Obs. 4 & 0.20$\dagger$ & 4.42$_{-0.03}^{+0.03}$ &  11.48 & 12.32$_{-0.84}^{+0.89}$ & 15.25$_{-1.02}^{+1.08}$ & {112.48/80}\\
Obs. 5 & - & - & - & - & 26.94$_{-2.17}^{+2.57}$  &  {86.84/78}\\
\hline
\end{tabular}
\label{tab 3}
\begin{flushleft}
$\dagger$ Error values are insignificant 
\end{flushleft}
\end{table*}

\subsection{Wide-band Observational Analysis}
\label{sec 4.3}
 {The intermittent re-brightenings, exhibited by GRS 1915+105 during the low-luminosity period, were vividly observed in the soft energies ($0.7 - 12$ keV) by \textit{NICER}. The spectral and timing properties of the source during each re-brightening have already been discussed in \S \ref{sec 4.2}. }
In this section, we constrain the broadband spectral and timing properties of the source by analyzing the simultaneous observations by \textit{AstroSat, NICER} and \textit{NuSTAR} (18 Epochs in Table \ref{tab 1}). The 18 wide-band observations are divided into two categories based on the X-ray activity of the source: The Quiet Phase - when the source exhibits steady and low X-ray flux (Epochs 6 – 9, 11, 12, 17 and 18 of Figure \ref{fig 2A}) and the Active Phase - when the source exhibits X-ray activities thereby producing an enhanced flux (Epochs 1 – 5, 10 and 13 -- 16 of Figure \ref{fig 2A}). The source exhibited high X-ray flux during Epoch 5($\sim58649$ MJD).  {However, due to the lack of \textit{NICER} observations between the period MJD 58636 and MJD 58656, we classify Epoch 5 under the active phase.} The PDS and energy spectra corresponding to all the wide-band observations are modeled and analyzed as mentioned in \S \ref{sec 3.1} and \S \ref{sec 3.2}. The spectral and timing properties are presented in the subsequent sections.

\subsubsection{Quiet Phase (QP)}
\label{sec 4.3.1}

The light curves corresponding to Epochs 6 -- 9, 11, 12, 17 and 18 showed no structured variability.  
The average \textit{MAXI} flux was $\sim0.15$ ph cm$^{-2}$ s$^{-1}$ (panel \textit{a} of Figure \ref{fig 2A}). The HR values are relatively higher (HR > 1, panel \textit{e} of Figure \ref{fig 2A}). The avg. count rate for every Epoch is mentioned in Table \ref{tab 1}. The power spectra obtained from all the Epochs are consistent with a power law model. None of these Epochs showed any indication of QPO features. The total rms varied between 7.1$_{-0.9}^{+0.6}$ to 22.2$_{+1.4}^{-1.2}$\% in $0.003 - 1$ Hz frequency range. The timing properties corresponding to all Epochs in the QP are summarized in Tables \ref{tab 5}.

In  Figure \ref{fig 6}, we show the spectra corresponding to the QP (spectra plotted in red) obtained from Epoch 18. 
A good fit for all the energy spectra is obtained using  Model-2.  The $\Gamma$ varied between 1.13$_{-0.06}^{+0.06}$ -- 1.73$_{-0.02}^{+0.02}$, while the kT$_{\rm e}$ ranged from 6.3$_{-0.2}^{+0.1}$ to 18.2$_{-3.7}^{+4.8}$ keV. All the Epochs showed obscuration with the N$_{\rm H_{\rm 1}}$ highly varying between $21_{-1}^{+1} - 545_{-72}^{+85} \times 10^{22}$ atoms cm$^{-2}$. The bolometric luminosities (L$_{\rm bol}$) during these Epochs varied between 0.001 L$_{\rm Edd}$ -- 0.004 L$_{\rm Edd}$. The fit values obtained for each of the Epochs are presented in Table \ref{tab 6}.

\subsubsection{Active Phase (AP)}
\label{sec 4.3.2}

The X-ray light curves obtained from Epochs 1 -- 5, 10 and $13 - 16$ showed high X-ray activity, when compared to the Epochs corresponding to the Quiet Phase. The \textit{MAXI} flux during these Epochs varied from 0.2 -- 0.6 ph cm$^{-2}$ s$^{-1}$ (panel \textit{a} of Figure \ref{fig 2A}). Each of these Epochs marks an event exhibited by the source during the three-year observation period. 
Epoch 1 corresponds to the decay phase of the major outburst with a flux value of 0.4 ph cm$^{-2}$ s$^{-1}$. During Epochs 2, 3 and 4, the source exhibited RB$_{\rm I}$ and the flux oscillated between 0.1 -- 0.4 ph cm$^{-2}$ s$^{-1}$. Epoch 10 corresponds to the rising phase of RB$_{\rm V}$, where the source exhibited $\rho^\prime$ variability class (see \S \ref{sec 4.2.3}). The \textit{LAXPC} light curve and CCD corresponding to Epoch 10 is shown in Figure \ref{fig 7} (panels \textit{a} and \textit{b} respectively), with the flux during each flare varying from $\sim150 - 250$ cts s$^{-1}$.  {The HR1  in the CCD is the ratio of count rates in $6 - 15$ keV and $3 - 6$ keV, while HR2 is the ratio of counts in $15 - 60$ keV  and $3 -6$ keV.} Epochs 13 -- 16 observe the source activity during the rise and the decay phase of RB$_{\rm VI}$, with the \textit{MAXI} flux at $\sim0.5$ ph cm$^{-2}$ s$^{-1}$. The HR value corresponding to the rebrightening phases (Epochs 10, 13 -- 16) showed lower HR values (HR < 0.4, panel \textit{e} of Figure \ref{fig 2A}).

\begin{figure} 
  \centering
  \hspace*{-1cm}
      \includegraphics[scale=0.3, angle=270]{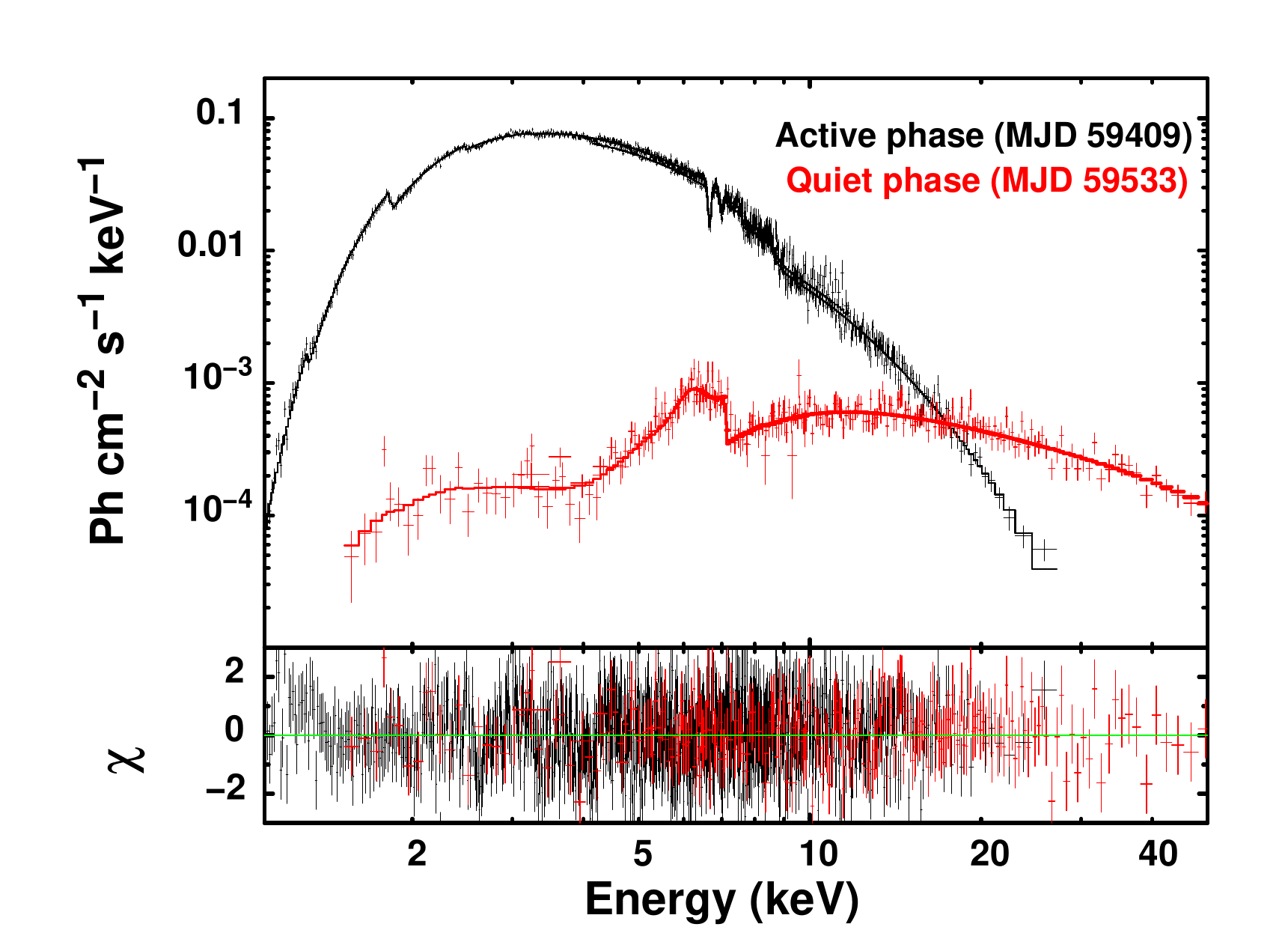}
     \label{fig 6}
  \caption{\label{fig 6} The broadband energy spectra corresponding to the Active phase (Epoch 13 - plotted in black) and the Quiet phase (Epoch 18 - plotted in red) obtained from simultaneous \textit{NICER} and \textit{NuSTAR} observations are shown here.  The source spectrum from the QP is fitted using Model-2, while the energy spectrum from the Active phase is fitted using Model-1. See text for details.}
\end{figure}

The PDS corresponding to Epoch 1 shows QPO at 2.08 Hz with a rms$_{\rm QPO}$ of $\sim12$\%. The power spectrum has a flat-top noise with a total rms of $\sim21.5$\%. Epochs 2, 3 and 4, pertaining to RB$_{\rm I}$, show high variability with the rms$_{\rm Tot}$ > 22\% and does not show any QPO signatures. Epochs 10 and 13 -- 16, pertaining to RB$_{\rm V}$ and RB$_{\rm VI}$ respectively, show QPO signatures with $\nu_{\rm QPO}$ varying from 20 to 200 mHz and rms$_{\rm QPO}$ varying between 8.7$_{-0.6}^{+0.6}$ -- 20.9$_{-1.0}^{+1.0}$\%. The total rms variability was relatively lesser with values varying from 12.1$_{-0.3}^{+0.3}$ -- 22.0$_{-1.2}^{+1.0}$\%. Panel \textit{c} of Figure \ref{fig 7} shows the PDS corresponding to Epoch 10 with a $\nu_{\rm QPO}$ detected at 23 mHz having a Q-factor of 6.3$_{-0.8}^{+0.8}$ and a rms$_{\rm QPO}$ of 8.7$_{-0.6}^{+0.5}$\%. The details of the timing properties obtained from the fits are mentioned in Table \ref{tab 5}.

The broadband source spectra obtained from all the Epochs in the AP are modeled using Model-1. The spectrum (plotted in black) in Figure \ref{fig 6} obtained from Epoch 13, demonstrates the typical broadband spectra corresponding to the AP. During the Epochs 1 -- 5,  the source exhibited the Low/Hard spectral nature with the $\Gamma$ and kT$_{\rm in}$ varying between 1.13$_{-0.01}^{+0.01}$ -- 1.73$_{-0.02}^{+0.02}$ and 0.25$_{-0.02}^{+0.01}$ -- 1.32$_{-0.04}^{+0.05}$ keV respectively. kT$_{\rm e}$ varied from 8.2$_{-0.2}^{+0.2}$ -- 16.6$_{-1.4}^{+1.7}$ keV. During the Epochs 10, 13 -- 16, we observe the source exhibiting softer spectral states, with $\Gamma$ varying from 1.8$_{-0.1}^{+0.1}$ to 2.8$_{-0.1}^{+0.1}$. kT$_{\rm in}$ and kT$_{\rm e}$ varied from 0.97$_{-0.01}^{+0.01}$ keV -- 1.59$_{-0.02}^{+0.01}$ keV and  3.1$_{-0.1}^{+0.1}$ to 12.9$_{-0.5}^{+0.4}$ keV, respectively. The best fitted model parameters are tabulated in Table \ref{tab 6}.

\begin{table*}
\centering
\caption{Details of model fitted parameters of PDS obtained from {\it LAXPC-AstroSat} (3 -- 60 keV), \textit{NICER} ($0.5 - 12$ keV) and \textit{NuSTAR} (3 -- 60 keV) observations of GRS~1915$+$105 for each Epoch corresponding to the wide-band observations (see Table \ref{tab 1}). Multiple \textit{Lorentzians} were used to fit the PDS.  Here, $\nu_{\rm QPO}$ represents the QPO frequency in Hz, $\sigma$ denotes the QPO significance and the total rms (rms$_{\rm Tot}$), given in percentage, is estimated for 0.003 -- 1 Hz frequency range. Errors for all the parameters are calculated with 90\% confidence.}
\begin{tabular}{|lcccccc|}
\hline
 Epoch & $\nu_{\rm QPO}$ & Q-factor & $\sigma$ & rms$_{\rm QPO}$ &  rms$_{\rm Tot}$ & $\chi^{2}/d.o.f$\\
 &(Hz)& & & (\%)& (\%)  & \\
\hline
 \rule{0pt}{10pt}1 & 2.08$_{-0.01}^{+0.01}$ & 10.71$_{-1.92}^{+2.08}$ & 4.79 & 11.94$_{-0.62}^{+0.58}$ & 21.38$_{-0.93}^{+1.13}$  &  {281.06/260}\\
 2 & - & - & - & - & 22.99$_{-2.06}^{+2.01}$  &  {139.17/141} \\
 3 & - & - & - & - & 22.9$_{-1.64}^{+1.80}$  &  {310.58/260}\\
 4 & - & - & - & - & 21.01$_{-2.33}^{+1.89}$ &  {116.5/111} \\
 5 & - & - & - & - & 7.50$_{-1.05}^{+0.99}$ &   {196.21/183}\\
 6 & - & - & - & - & 14.18$_{-0.99}^{+1.04}$ &  {123.5/103}\\
 7 & - & - & - & - & 22.18$_{-1.26}^{+1.37}$  &  {143.1/123}\\
 8 & - & - & - & - & 12.38$_{-0.94}^{+0.94}$  &  {107.1/109}\\
 9 & - & - & - & - & 7.05$_{-0.93}^{+0.61}$  &  {92.38/82}\\
 10 & 0.02$\dagger$ & 6.31$_{-0.80}^{+0.79}$ & 13.63 & 8.70$_{-0.63}^{+0.51}$ & 12.08$_{-0.30}^{+0.38}$  &  {139.58/105}\\
 11 & - & - & - & - & 10.79$_{-0.41}^{+0.48}$ &   {105.7/84}\\
 12 & - & - & - & - & 5.91$_{-0.36}^{+0.31}$  &  {101.6/93}\\
 13 & 0.18$\dagger$ & 3.15$_{-0.31}^{+0.34}$ & 19.11 & 15.93$_{-2.23}^{+1.97}$ & 20.39$_{-1.15}^{+1.28}$  &  {104.1/106}\\
 14 & 0.15$\dagger$ & 4.89$_{-0.31}^{+0.40}$ & 7.67 & 12.57$_{-1.88}^{+1.57}$ & 18.06$_{-0.68}^{+0.66}$  &  {100.9/84}\\
 15 & 0.17$\dagger$ & 2.71$_{-0.29}^{+0.51}$ & - & 13.62$_{-0.84}^{+0.91}$ & 18.57$_{-1.09}^{+1.02}$  &  {119.4/84}\\
 16 & 0.18$\dagger$ & 3.78$_{-0.64}^{+0.55}$ & - & 20.90$_{-0.96}^{+1.00}$ & 22.01$_{-1.20}^{+1.04}$  &  {123.2/89}\\
 17 & - & - & - & - & 13.13$_{-1.37}^{+1.24}$  &  {98.5/89}\\
 18 & - & - & - & - & 15.60$_{-1.20}^{+1.46}$  &  {112.6/107}\\
\hline
\end{tabular}
\label{tab 5}
\begin{flushleft}
      $\dagger$ Error values are insignificant 
\end{flushleft}
\end{table*}

\begin{figure}
\centering
\hspace*{-0.9cm}
\includegraphics[scale=0.305, angle=270]{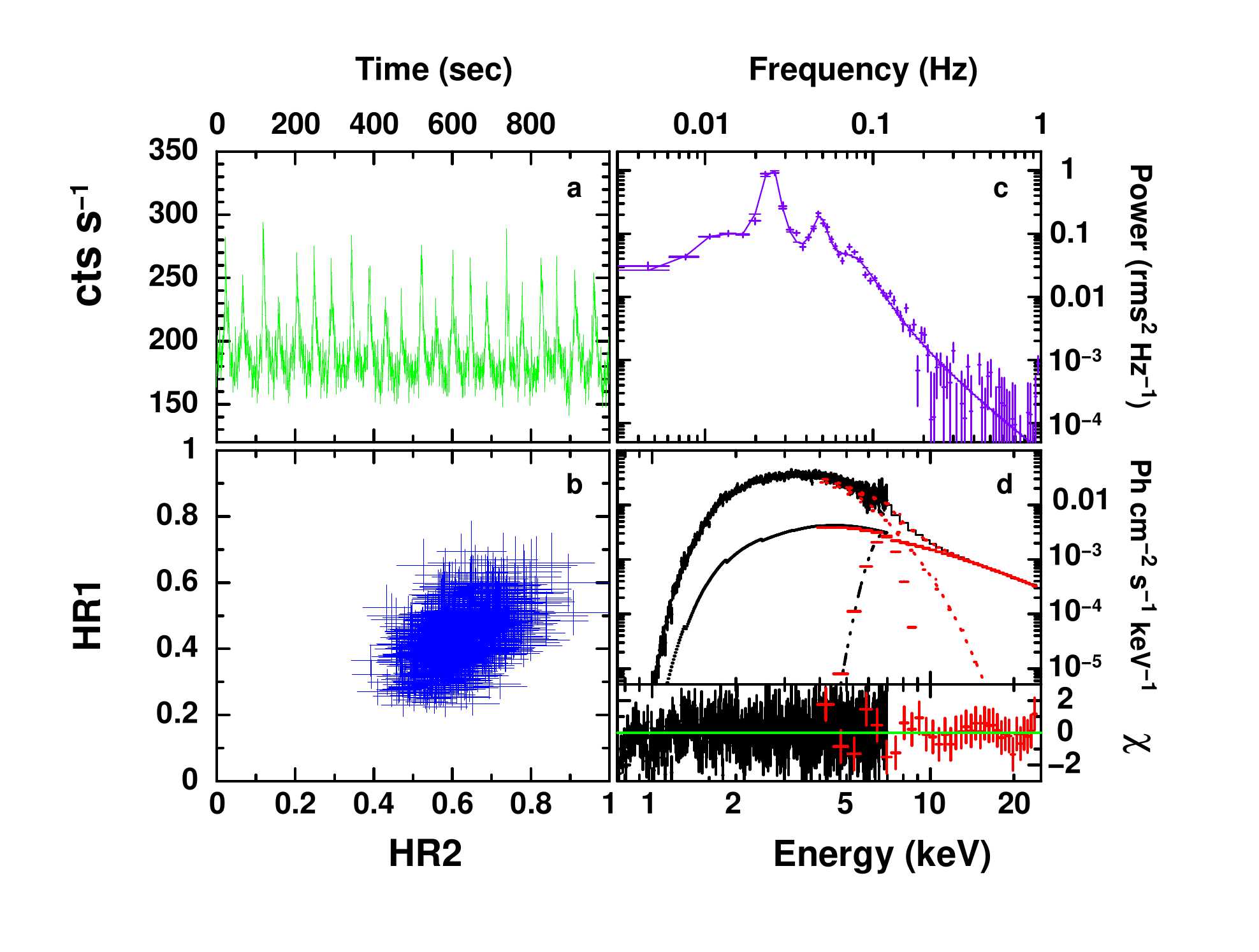}
\vspace*{-0.65cm}
\caption{ {Background subtracted 1~s binned \textit{LAXPC} light curve (3 -- 60 keV) corresponding to Epoch 10 (Active phase) is shown in panel \textit{a}. The CCD of the source  is shown in panel \textit{b}, where HR1 is plotted on the X-axis and HR2 is plotted on the Y-axis. Panel \textit{c} shows the PDS generated from the 0.01 s binned light curve in 3 -- 60 keV energy band with a QPO at 23 mHz and a Q-factor of $6.31\pm0.8$. Panel \textit{d} depicts the broadband energy spectrum from the \textit{SXT} and \textit{LAXPC} for the energy range 0.7 -- 25 keV. The \textit{SXT} and \textit{LAXPC} data are plotted in black and red colours, respectively.  The spectrum is best fitted using Model-1. The dotted lines represents the disc and the Comptonizing model components used to fit the data, while the dashed curve shows the gaussian component used to fit the Fe emission line.} \label{fig 7}}
\end{figure}

\subsection{Absorption and Emission Features}
\label{sec 4.4}

\begin{figure}
\centering
\hspace*{-0.5cm}\includegraphics[scale=0.34, angle=270]{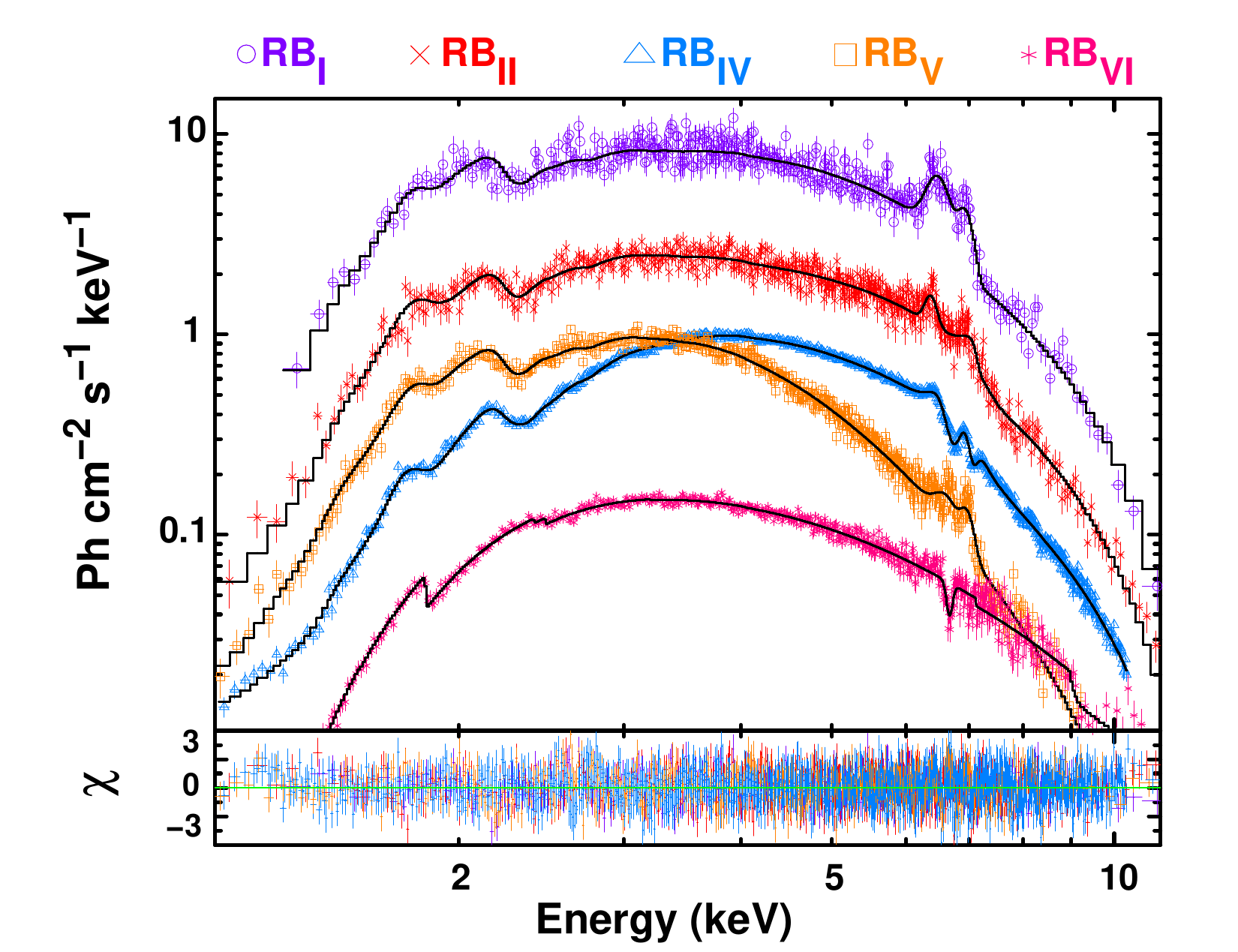}
\caption{An overplot of the energy spectra obtained from  {RB${_{\rm I}}, $RB${_{\rm II}}$, RB${_{\rm IV}}$, RB${_{\rm V}}$ and RB${_{\rm VI}}$} is presented here in purple, red, orange, sapphire and magenta colored plots respectively. The \textit{NICER} energy spectra corresponding to RB${_{\rm I}}$ (Obs. 4), RB${_{\rm II}}$ (Obs. 1), RB${_{\rm V}}$ (Obs. 2)  show the presence of two Fe emission lines between $6 - 7$ keV. The spectra associated with RB${_{\rm IV}}$ (Obs. 1) show the presence of two gaussian absorption lines, while  RB${_{\rm VI}}$ (Obs. 4) shows one gaussian absorption line. \label{fig 8}}
\end{figure}

The characteristic features in the X-ray spectrum include the prominent emission and absorption lines superposed on the spectral continuum. The emission lines are essentially the fluorescent line photons from the disc, originated from the illumination of the disc by the hard X-ray photons \citep{1989MNRAS.238..729F}. The absorption by the outflowing plasma from the accretion disc generates absorption lines \citep{2009Natur.458..481N}. All the observations considered in our work (wide-band observations (see Table \ref{tab 1}) and individual \textit{NICER} observations (see Table \ref{tab 2}) also showed the presence of prominent Fe absorption and emission line features  {(see Figure \ref{fig 8})}. We use the \textit{gaussian} and the \textit{gabs} models to estimate the features of the emission and the absorption lines respectively (as already described in \S \ref{sec 3.2}). Broad and narrow emission lines were detected between the energy ranges 6.4 -- 8.3 keV (see Tables \ref{tab 4} and \ref{tab 6}). The centroid energies of these lines correspond to the neutral Fe K$\alpha$, Fe XXV K$\alpha$, Fe XXVI K$\alpha$, and the Ni XXVIII K$\alpha$ energies at 6.4 keV, 6.7 keV, 6.97 keV, and 8.10 keV respectively. The strength of these lines are measured in terms of the Equivalent Width (EW), which is estimated using the standard definition,
\begin{equation} \label{eqn1} EW = \int_{E_{1}}^{E_{2}} \frac{F{_c}(E)-F(E)}{F{_c}(E)} dE,
	\end{equation}

where, $\rm F{_{\rm c}}(E)$ is the flux in the continuum, $\rm F(E)$ the flux in the line at Energy (\textit{E}). E$_1$ and E$_2$ represent the lower and upper energy limits of the observed line (see \citealt{2004fost.book.....S}). Emission lines were predominantly observed in the QP and re-brightening phases - RB${_{\rm I}}$, RB${_{\rm II}}$ and RB${_{\rm V}}$. The EW of the emission lines is observed to vary from 70 -- 990 eV.  {However, the source also exhibited broad emission lines where the EW varied between 1020 -- 3260 eV. Emission lines with EW $\geq~1$ keV is not a commonly observed feature in X-ray binaries, but instead is considered as a classic indicator of a Compton-thick (N$_{\rm H_{\rm 1}} \geq 10^{24}$ atoms cm$^{-2}$) obscuration generally seen in AGNs \citep{2006ApJ...648..111L}. A Compton thick obscuration suppresses the continuum beneath the neutral line, thus leading to an increase in the EW of the Fe K$\alpha$ line.} However, the absorption lines observed throughout the observation period were narrow with the EWs varying from 120 to 590 eV. These narrow absorption lines were observed during RB${_{\rm IV}}$ and RB${_{\rm VI}}$ in the energy ranges 6.4 keV -- 7 keV. The line properties obtained from the best-fit models are quoted in Tables \ref{tab 4} and \ref{tab 6}. The broad emission lines during the hard state and the narrow absorption lines during the relatively softer spectral states were formerly observed in GRS 1915+105 \citep{2009Natur.458..481N}. It is speculated that the broad emission lines originate when the inner accretion disc is illuminated by the hard X-ray photons from the jet/corona, while the narrow absorption lines  {are due} to the winds in the accretion disc.

The EW of the absorption lines enable us to estimate the column densities of  Fe XXV and Fe XXVI elements, using the relation, 
\begin{equation} \label{eqn2}
	W_{\lambda} = (\pi e^{2}/M_{e}c^{2})N_{j}\lambda^{2}f_{ij} = 8.85\times10^{-13}N_{j}\lambda^{2}f_{ij},
	\end{equation}

where, $W_{\lambda}$ is the EW of the line and $\lambda$ is the wavelength in centimeters, $f_{ij}$ is the oscillator strength and is equal to 0.798 and 0.416 for Fe XXV and Fe XXVI elements, respectively (with reference to \citealt{1978ppim.book.....S, 2002ApJ...567.1102L}). The energy of the lines (in keV) is converted to wavelength (in cm) using the relation, $E = hc/\lambda$, where \textit{h} is the plank's constant ($6.626\times 10^{-34}$ Js) and \textit{c} is the velocity of light ($3\times10^{10}$ cm/s). The ion column density ($N_{j}$) thus obtained helps in constraining the physical parameters of the absorbing plasma. Our estimates show the  {Fe column density values} to be varying between 10$^{16} - 10^{18}$ atoms cm$^{-2}$. These moderate values of ion column densities suggest the kinetic temperature of the absorbing plasma (kT$_{\rm Fe}$) to be $\geq 25$ keV \citep{2000ApJ...539..413K, 2001A&A...379..540S}. With reference to \citealt{2001PASJ...53..179Y}, if we assume the absorbing plasma to be in hydrodynamical equilibrium in the direction vertical to the plane, we can calculate the radius of the absorbing plasma from the center (r) using the relation suggested in \citealt{1973A&A....24..337S}, 
\begin{equation} \label{eqn3}
\left( \frac{h}{r} \right)^{2} \frac{GMm_{\rm H}}{r} \simeq kT_{\rm th},
	\end{equation}
	
where, $\frac{h}{r}$ = tan($90^{\circ}-i$), \textit{i} being the inclination angle ($60^{\circ}$; \citealt{2014ApJ...796....2R}), G is the gravitational constant, M is the mass of the black hole and m$_{\rm H}$ is the mass of the hydrogen atom. kT$_{\rm th}$ is the thermal temperature and can be estimated using the relation, $\rm kT_{\rm th} = \left(m_{\rm H}/m_{\rm Fe} \right) kT_{Fe}$, where m$_{\rm H}$ is the mass of the Fe atom \citep{2001PASJ...53..179Y}. For kT$_{\rm Fe}\geq25$ keV, the absorbing plasma is found at a distance r $\leq 2\times10^{10}$ cm.  {\citealt{2020ApJ...904...30M} estimated the radius of the inner hot absorption zone, from where the winds are launched, to be at r < $10^9$ cm.} 

\section{Discussion}
\label{sec 5}

We performed a comprehensive study of the spectral and timing characteristics of the source during the low-luminosity `obscured' phase between March 2019 --  {November 2021}. During this period, the source is seen exhibiting a low-luminosity phase at length, along with the occasional re-brightening phases. Below, we present a cohesive explanation of the overall evolution of the source properties during each of these phases.

\subsection{State Transitions during the Re-brightening Phases}
\label{sec 5.1}

\begin{figure*}
\centering
\hspace*{-0.2cm}\includegraphics[scale=0.45]{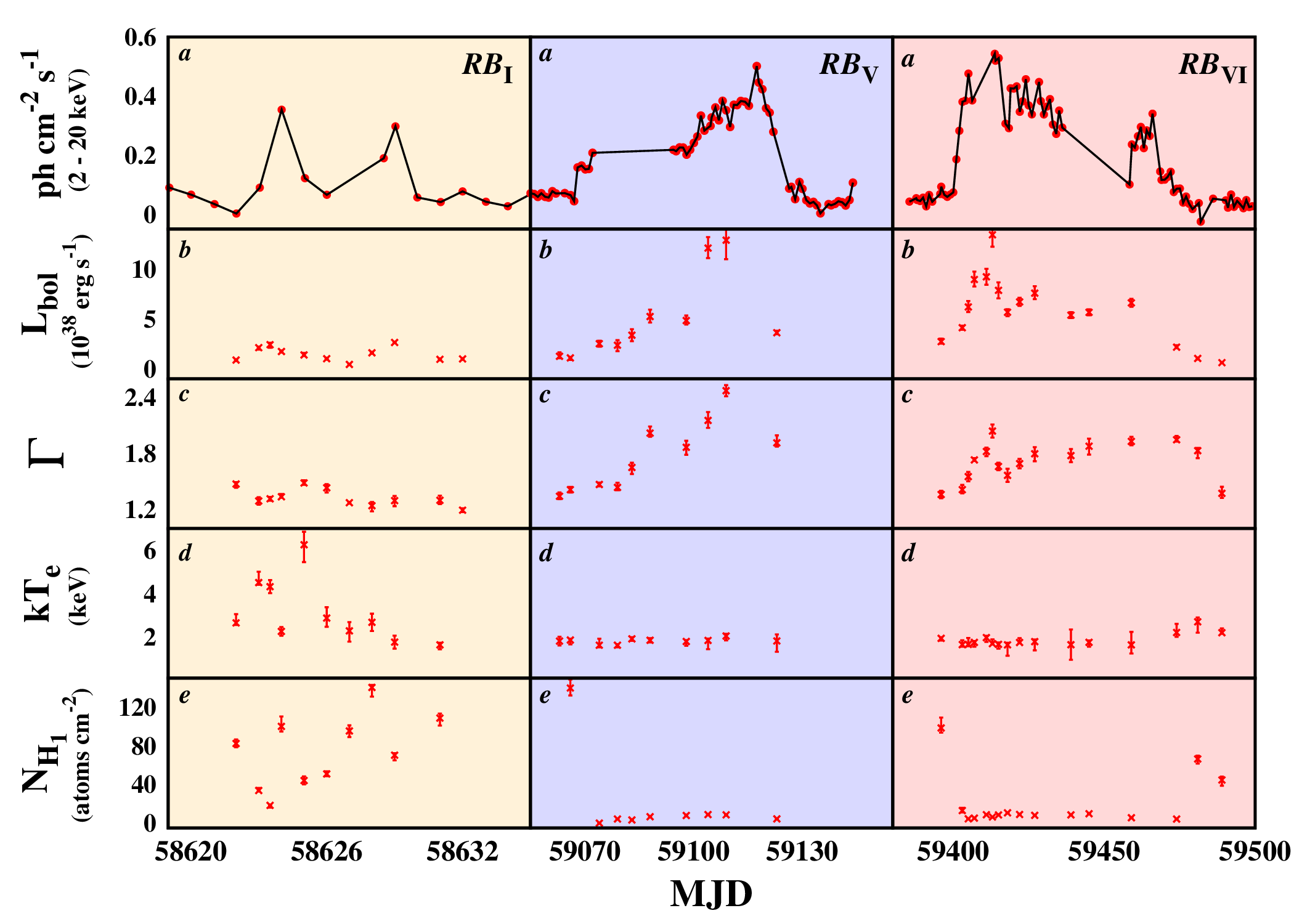}
\caption{ {Overall time evolution of some important spectral features with time during RB$_{\rm I}$, RB$_{\rm V}$ and RB$_{\rm VI}$ is shown in yellow, blue and red shaded columns respectively. The top most panel (panel \textit{a} in every column) shows the \textit{MAXI} flux of the source in $2-20$ keV energy range, during each re-brightening. Panels \textit{b}, \textit{c}, \textit{d} and \textit{e} show the evolution of the bolometric Luminosity (L$_{\rm bol}$ in 10$^{38}$ erg s$^{-1}$), $\Gamma$, kT$_{\rm e}$ (keV) and the additional hydrogen column density (atoms cm$^{-2}$), respectively.} \label{fig 9}}
\end{figure*}

GRS 1915+105 exhibited 6 major re-brightenings - RB$_{\rm I}$, RB$_{\rm II}$, RB$_{\rm III}$, RB$_{\rm IV}$, RB$_{\rm V}$ and RB$_{\rm VI}$ (see  panel \textit{a} of Figure \ref{fig 2A}) throughout the three-year observation period.  {RB$_{\rm I}$ was a series of flares,  RB$_{\rm II}$ and RB$_{\rm IV}$ quick flares (spanning for a few ksec) and RB$_{\rm V}$ and RB$_{\rm VI}$ were prolonged re-brightenings (spanning for $\sim100$ days and $\sim150$ days respectively). Figure \ref{fig 9} shows the overall evolution of few important spectral parameters like, L$_{\rm bol}$ (in 10$^{38}$ erg s$^{-1}$),  $\Gamma$, kT$_{\rm e}$ (in keV) and N$_{\rm H_{1}}$ (atoms cm$^{-2}$) during the rebrightening phases - RB$_{\rm I}$, RB$_{\rm V}$ and RB$_{\rm VI}$.  {Figure \ref{fig 9} also includes the results from the analysis of the additional 40 \textit{NICER} observations which are not tabulated in Table \ref{tab 2}, as stated in \S\ref{sec 2.3}.}} 
The source exhibited state transitions during the prolonged re-brightenings RB$_{\rm V}$ (Figure \ref{fig 4}) and RB$_{\rm VI}$ (Figure \ref{fig 5}).  {At the beginning of both RB$_{\rm V}$ and RB$_{\rm VI}$, the source is detected in the hard spectral state during the low phase (Obs. 1), with $\Gamma$ of $\sim1.3$ (see panel \textit{c} corresponding to RB$_{\rm V}$ and RB$_{\rm VI}$ in Figure \ref{fig 9}) and an almost constant electron temperature, kT$_{\rm e}\sim 2$ keV (see \S \ref{sec 4.2.3} and \S \ref{sec 4.2.4}). As the source progresses into the rise and the peak phase of these re-brightenings, $\Gamma$ is seen to increase from 1.34$_{-0.04}^{+0.03}$ to 2.47$_{-0.06}^{+0.03}$ during RB$_{\rm V}$ and 1.36$_{-0.02}^{+0.02}$ to 2.04$_{-0.04}^{+0.02}$ during RB$_{\rm VI}$. Source exhibited a maximum luminosity  L$_{\rm bol}$ of $12.8\times10^{38}$ erg s$^{-1}$ and $13.4\times10^{38}$ erg s$^{-1}$  during the peak of RB$_{\rm V}$ and RB$_{\rm VI}$, respectively (see panel \textit{b} in the blue and red shaded regions in Figure \ref{fig 9}). The decay phase is characterized by a decrease in $\Gamma$, to 1.91$_{-0.01}^{+0.03}$ and 1.69$_{-0.05}^{+0.02}$ during the decay phases (Obs. 4) of RB$_{\rm V}$ and RB$_{\rm VI}$, respectively. A further decrease in the photon index ($\Gamma\sim1.3$) is observed as the source descends to the low phase after the decay (Obs. 5 in RB$_{\rm VI}$), indicating the hard spectral nature of the source. However, kT$_{\rm e}$ (see panel \textit{d} in the blue and red shaded regions in Figure \ref{fig 9}) is found to remain constant throughout the re-brightening phases. The rise, peak and the decay phases (Obs. 2, 3 and 4 respectively) are recognized as the intermediate/soft state. The total rms variability also decreases as the source progresses from the hard to the soft state (13.1$_{-0.8}^{+1.1}$\% to 8$_{-0.8}^{+0.9}$\% during RB$_{\rm V}$, 32.0$_{-3.2}^{+2.9}$\% to 17$_{-1.4}^{+1.3}$\% during RB$_{\rm VI}$). }

Similar evolution pattern in the spectral and timing properties has already been observed during the 2018 mini-outburst of MAXI J1535-571 \citep{2020MNRAS.496.1001C} and 2017 mini-outburst of GRS 1739-278 \citep{2017MNRAS.470.4298Y}. Several other LMXBs also have previously exhibited re-brightenings \citep{2012MNRAS.423.3308J, 2017MNRAS.470.4298Y, 2020MNRAS.496.1001C}. However, spectral state transitions are not witnessed during every re-brightening. 
Only 2 BH LMXBs - MAXI J1535-571 \citep{2020MNRAS.496.1001C} and GRS 1739-278 \citep{2017MNRAS.470.4298Y}, so far, have exhibited spectral state transitions from the hard state to the soft state during the re-brightenings. However, a disc component is seen in both sources during the peak of the outburst with 2.1 < $\Gamma$ < 2.7 and 0.33 < kT$_{\rm in}$ (keV) < 0.5, whereas these source showed no indication of the disc component during the hard state, with 1.5 <  $\Gamma$ < 2. These two sources also showed hysteresis in the HID, thereby generating a q-track in the HID. Nevertheless, we not observe a q-track in the HID during RB$_{\rm V}$ and RB$_{\rm VI}$. Based on an analogy of the spectral and timing characteristics of GRS 1915+105 with MAXI J1535-571 and GRS 1739-278, it can be concluded that the source undergoes the following sequence of transitions: hard $\rightarrow$ intermediate $\rightarrow$ soft state $\rightarrow$ intermediate $\rightarrow$ hard state (during RB$_{\rm V}$) and  hard $\rightarrow$ soft state $\rightarrow$ hard state (during RB$_{\rm VI}$).

All the mini-outbursts/re-brightenings in MAXI J1535--571, GRS 1739--278 and GRS 1915+105 are seen to have completely different timescales. In case of MAXI J1535-571, re-flares occurred soon after the major outburst, whereas in GRS 1739--278, the time gap between the major outburst and the mini-outburst is not clearly understood due to the observational gap. However, \cite{2020MNRAS.496.1001C} predicts a time gap of < 200 days between the major and the mini-outburst. But, in the case of GRS 1915+105, RB$_{\rm V}$ happened 500 days after the major outburst and RB$_{\rm VI}$  is seen occurring 200 days after the decay of RB$_{\rm V}$. The periodicity of the mini-outbursts in MAXI J1535--571 and GRS 1739--278 was estimated to vary between $\sim$20 -- 35 days. In GRS 1915+105, RB$_{\rm V}$ and RB$_{\rm V}$ lasted for $\sim100$ days and $\sim150$ days, respectively. A comparison of the timescales seen in all the three sources does not lead us to the common cause that triggers these re-brightenings. There exists several models in the literature that explains the origin of the re-brightenings \citep{1993ApJ...408L...5C, 1993A&A...279L..13A, 1998MNRAS.293L..42K, 2005ApJ...625..923S}. But based on the results, we postulate that  the re-brightenings/mini-outbursts are small-scale outbursts  \citep{2016ApJ...817..100P}, where mini-outbursts develop and progress in a way similar to the main outburst. The instability is assumed to be  triggered at some location in the outer disc, which gradually increases the disc density and temperature. The mass accretion rate increases as this instability advances and propagates inwards as a heating wave, thus causing a re-brightening or a mini-outburst. The detection of the disc component in GRS 1915+105 as the spectral state of the source softens towards the peak, complies with the above scenario.

\subsection{Variability during the Re-brightenings}
\label{sec 5.2}

GRS 1915+105 has exhibited 15 variability classes since its discovery. During the major outburst, the average count rate exhibited by the source, during each of the variability classes, varied from $1 - 50$ kcts s$^{-1}$ (\cite{1999ApJ...527..321M, 2000A&A...355..271B, 2001MNRAS.324..267N, 2002JApA...23..213N, 2022MNRAS.510.3019A, 2022MNRAS.512.2508M} and references therein). It was also reported that the limit-cycle oscillations seen during certain variability classes, disappeared at an average count rate < 5 kcts s$^{-1}$ \citep{2002ApJ...576..908J}. However, during the recent `obscured' phase, the source has displayed  $\rho, \lambda$ and $\delta$ variability classes during RB$_{\rm V}$ and RB$_{\rm VI}$ (Figures \ref{fig 4}a and \ref{fig 5}a) exhibiting an average count rates of $\sim30, 120$ and 250 cts s$^{-1}$, respectively.  {GRS 1915+105 also exhibited $\rho^\prime$ variability class (\S\ref{sec 4.2.4}), a variant of the typical $\rho$ class (see \citealt{2022MNRAS.510.3019A}) during the rise phase of RB$_{\rm V}$. Using the Modified Hindmarsh-Rose (MHR) model,  \cite{2020MNRAS.496.1697M} shows that $\rho^\prime$ could be a result of the slight modulation of time dependent input function \textit{(J(t))}. We also categorize the source to belong to $\chi$ variability class during the low-luminosity phase, where the source exhibited Low/Hard spectral state ($\Gamma \sim$ 1.13$_{-0.06}^{+0.06}$ -- 1.73$_{-0.02}^{+0.02}$ and  kT${_{\rm e}} \sim$ 6.3$_{-0.2}^{+0.1}$ -- 18.2$_{-3.7}^{+4.8}$ keV, see \S \ref{sec 4.3.1}). However, the PDS did not explicitly show any QPO features during the the hard state, which could be a result of low statistics. 
The source also exhibits a sequential class transition from $\chi \rightarrow \lambda \rightarrow \rho$, during the rising phase of RB$_{\rm V}$. The time evolution of the variability classes in GRS 1915+105 depicted by the MHR model \citep{2020MNRAS.496.1697M} indicate that the source makes transition from stable states - states showing stable equilibrium patterns (classes $\phi, \chi, \alpha^{\prime\prime}, \theta, \xi$ and $\omega$) to an unstable state - state showing unstable equilibrium pattern ($\rho$ class) via transition segment ($\delta, \gamma, \lambda, \kappa$ and $\alpha^{\prime}$ classes), as the time dependent input function \textit{(J(t))} varies. This MHR model is in complete accordance with the sequential transition from $\chi \rightarrow \lambda \rightarrow \rho$, ~exhibited by the source during RB$_{\rm V}$.}

The variability classes is a unique feature particular to this source. These unique variability classes depict the accretion and ejection limit cycles in an unstable disc \citep{1997ApJ...488L.109B, 2011ApJ...737...69N}. Based on the statistical analysis of \textit{RXTE} observations of the source between 1996 April to 2000 May, it was inferred that the unique variability classes occurred when the source radiated at exorbitant luminosities and satisfied the criteria L/L$_{\rm Edd} \geq 1$ \citep{2004MNRAS.349..393D}. However, recently, GRS 1915+105 began exhibiting variability classes at L/L$_{\rm Edd} \leq 1$ \citep{2020MNRAS.499.5891S, 2022MNRAS.510.3019A}. In hindsight, this decrease in the source luminosities were anticipated, because of the depletion of the matter in the accretion disc with the persistent X-ray activity. Similarly, IGR J17091--3624 \citep{2011ApJ...742L..17A, 2015ApJ...807..108I, 2017MNRAS.468.4748C, 2018Ap&SS.363..189D, 2021MNRAS.501.6123K} also exhibited these variability patterns at $\sim$20 -- 30 \% L$_{\rm Edd}$, thereby manifesting that the source need not necessarily radiate at  luminosities $\geq$ L$_{\rm Edd}$, in order to exhibit unique variability classes.

In addition to the observations, the results from the time-dependent disc models and the simulations predict the outset of limit-cycle instabilities at L/L$_{\rm Edd} \geq 0.3$ \citep{2000ApJ...535..798N, 2002ApJ...576..908J}. Nonetheless, the above mentioned calculated predictions stand questionable based on the recent activities of GRS 1915+105, where the source exhibited variability classes at unabsorbed luminosities varying between 0.01 -- 0.004 L$_{\rm Edd}$, which is $\ll$ 0.3 L$_{\rm Edd}$. At these luminosities, LMXBs generally show least activity/variabilities. With the discrepancies in the disc-instability models to explain the limit-cycle oscillation behaviour, magnetic fields can be considered as an alternative way to explain the limit-cycle instabilities. The lack of threshold large-scale magnetic fields of uniform polarity eventually fails in sustaining the thermal stability in the accretion disc \citep{2016MNRAS.462..960S}. This could be a cause for the peculiar variabilities seen in GRS 1915+105. However, the key questions on how the phenomena that triggers the unique variability patterns in the system is invoked even at extremely low luminosities (L $\sim$ 0.01 L${_{\rm Edd}}$), is yet to be understood.

\subsection{Dynamical Obscuration in the System}
\label{sec 5.3}

An incessant presence of `obscuration' in GRS 1915+105 since May 2019 has been detected from the spectral analysis. This obscuration is detected to be highly variable and in-homogeneous, i.e. PCF < 1 (see \citealt{2017MNRAS.471.1797M}). The column density due to the local obscuration (N$_{\rm H_{\rm 1}}$) and the partial covering fraction (PCF) varied drastically within a few minutes.

During the prolonged low-luminosity phase (\S \ref{sec 4.3.1}), the source exhibited non-deviant Low/Hard spectral state properties. However, the obscuration in the system was highly dynamic and random with N$_{\rm H_{\rm 1}}$ varying between $20 - 550\times 10^{22}$ atoms cm$^{-2}$ and the PCF varied between 0.38 -- 0.96 (see Table \ref{tab 6}).  {In case of the prolonged re-brightening phases (RB$_{\rm V}$ and RB$_{\rm VI}$), the source showed intrinsically evolving spectral characteristics as well as a methodically evolving local obscuration medium (see panel \textit{a} in the blue and red shaded columns in Figure \ref{fig 9}).}  {The obscuration varied in a pattern, where the least column density (N$_{\rm H_{\rm 1}}$) was detected during the active phase of the re-brightening (rise/peak/decay phases of the re-brightenings) and a maximum density was detected in the low phase (hard state) either before the rise or after the decay of the re-brightenings. A minimum N$_{\rm H_{\rm 1}}$ value of $4\times 10^{22}$ atoms cm$^{-2}$ was observed near the decay phase of RB$_{\rm V}$ and the peak phases of RB$_{\rm VI}$.}  {However, an increased N$_{\rm H_{\rm 1}}$ value of $\sim150\times10^{22}$ atoms cm$^{-2}$ and $\sim100\times10^{22}$ atoms cm$^{-2}$ was observed during the low phase/hard state of both RB$_{\rm V}$ and RB$_{\rm VI}$, respectively. The PCF varied randomly between 0.6 -- 0.79 and 0.37 -- 0.93 throughout RB$_{\rm V}$ and RB$_{\rm VI}$, respectively.} 
 {This varying trend in N$_{\rm H_{\rm 1}}$ is also observed during RB$_{\rm I}$ (see panel \textit{a} corresponding to RB$_{\rm I}$ in Figure \ref{fig 9}), where N$_{\rm H_{\rm 1}}$ decreased significantly to $\sim18\times10^{22}$ atoms cm$^{-2}$ during the flare in comparison with the other phases of RB$_{\rm I}$  (see Table \ref{tab 4}).}

An identical scenario was observed in V404 Cyg during its outburst in 2015. The \textit{Swift} observations of the source revealed fast and highly variable obscuration (N$_{\rm H_{\rm 1}}$ $\sim10^{21} - 10^{24}$ atoms cm$^{-2}$) within the system \citep{2017MNRAS.471.1797M}. The authors suggest a clumpy Compton thick outflow to explain the fast variable obscuration in the system (see Figure 10 in \citealt{2017MNRAS.471.1797M}).  {This clumpy Compton thick  outflow aptly describes the nature of the obscuring material in GRS 1915+105. The high EW of the neutral Fe K$\alpha$ line (EW $\geq$ 1 keV) obtained from the analysis of GRS 1915+105 (see \S \ref{sec 4.4}), can also be considered a proponent indicator of the presence of a Compton thick absorbing material \citep{1998A&A...338..781M}.  The} 
 {decrease in the density of the obscuring medium during the flares is seen in both GRS 1915+105 and V404 Cyg.} \cite{2017MNRAS.471.1797M} presumes that the radio flare preceding the X-ray flare in V404 Cyg drives the obscuring medium away, at least for a short span, which leads to a decreased density and the PCF during the flare. This explanation can be adapted to justify the decrease in N$_{\rm H_{\rm 1}}$ during the flare in RB$_{\rm I}$, as there exists a precursory radio flare to RB$_{\rm I}$. But, in case of RB$_{\rm V}$ and RB$_{\rm VI}$, we observe a decreased radio activity as the source moves towards the peak during RB$_{\rm V}$ and RB$_{\rm VI}$ (\textit{ALMA} observations as reported in \citealt{2021MNRAS.503..152M}). This indicates that there is some other mechanism in the system that drives away the obscuring medium. Although their results and our results show some similarities to some extent, we understand that this explanation cannot be considered appropriate  for GRS 1915+105 because both V404 Cyg and GRS 1915+105 differs fundamentally with regards to the accretion rate. V404 Cyg accreted at  {Eddington/super-Eddington} rates thereby inflating the inner accretion disc (slim disc; \citealt{1988ApJ...332..646A}), thus causing the clumpy and Compton thick outflows, whereas GRS 1915+105 accretes at  {sub-Eddington} rates which is inadequate to generate a clumpy outflow.

In the recent works, \cite{2020ApJ...904...30M} and \cite{2021ATel14792....1N} identified the obscuring medium as various layers of absorbing zones extending radially out. \cite{2020ApJ...904...30M} detected the winds emanating from the inner hot absorption zone, at a radius < $10^9$ cm and N$_{\rm H_{\rm 1}}$ $\sim10^{23}$ atoms cm$^{-2}$. These winds are elucidated as failed winds, which could not be launched to infinity due to the lack of magnetic field strength. The winds eventually surround the central engine thus obscuring the system. Alternatively, \cite{2021ATel14792....1N} deduced the obscuring medium at a distance $\sim10^{11}$ cm with a density $\sim10^{12}-10^{13}$ cm$^{-3}$. The radial and the vertical profile of the disc, from the deduced density and radius values, suggested an inflation of the outer disc. The inflated disc acted like a torus thereby partially or completely obscuring the inner accretion disc (see also \citealt{2014ApJ...786L..20M}). However, it is not viable to capture the dynamic nature of the obscuration with models as suggested in \cite{2020ApJ...904...30M, 2021ATel14792....1N, 2021MNRAS.503..152M} based on a limited sample of observations. In our study, we consider a large sample of observations to study the fast and the complex evolution of the obscuring medium. Yet, there is no consensus on the processes causing the obscuration and the dynamics of the obscuration. An insight on the dynamics of the obscuring medium may be obtained by a quantitative analysis of the source using a time-dependent model for obscuration, which is beyond the scope of the present work.

\section{Summary}
\label{sec 6}

In this paper, we performed a thorough and comprehensive spectral and timing analysis of \textit{AstroSat, NICER} and \textit{NuSTAR} observations pertaining to the low-luminosity `obscured' phase, that GRS 1915+105 has been exhibiting since May 2019. Based on the results, we present a cohesive summary of the evolution of the spectral and timing characteristics of the source.

\begin{itemize}
   \item GRS 1915+105 exhibited multiple re-brightenings both in X-ray and radio energies. The bolometric luminosities (L$_{\rm bol}$) of the source varied between 0.004 L$_{\rm Edd}$ at the low-luminosity phase to 0.01 L$_{\rm Edd}$ during the peak of the re-brightening phases.
    
    \item The source exhibited state transitions during the prolonged re-brightening phases. The source was tracked making a transition from hard $\rightarrow$ intermediate $\rightarrow$ soft state $\rightarrow$ intermediate $\rightarrow$ hard state, during RB$_{\rm V}$ and hard $\rightarrow$ soft state $\rightarrow$ hard state, during RB$_{\rm VI}$.
    \item  GRS 1915+105 displayed the characteristic variability classes, $\rho, \lambda$ and $\delta$ at unabsorbed luminosities of  0.01 L$_{\rm Edd}$, 0.02 L$_{\rm Edd}$ and 0.02 L$_{\rm Edd}$ respectively. Although QPOs could not be detected, based on the spectral characteristics the source is classified to belong to the $\chi$ variability class during the low-luminosity phases. 
    \item The source revealed multiple Fe absorption and emission line features between 6 -- 8 keV with EW varying between 70 eV -- 3.26 keV. Fe XXV and Fe XXVI ion column density varied between 10${^{16} - 10^{18}}$ atoms cm$^{-2}$. The distance of the absorbing plasma was constrained to be $\leq2\times10^{10}$ cm.
      \item The source exhibited a highly dynamic obscuration with column density varying between $\sim10^{22} - 10^{24} $atoms cm$^{-2}$ throughout the 3-year observation period.
\end{itemize}

\section*{Acknowledgements}

 {We thank the anonymous reviewer for his/her suggestions and comments that helped to improve the quality of this manuscript.} AMP, AN acknowledge the financial support of Indian Space Research Organisation (ISRO) under RESPOND program Sanction order No. DS-2B-13012(2)/19/2019-Sec.II. AMP acknowledges the PI of this project, Dr. Baishali Garai, for the relentless support and guidance.  {AMP also thanks Dr. Dominic Walton (University of Hertfordshire, Hatfield, UK) for his support in pursuing this work.} This publication uses data from the \textit{AstroSat} mission of the ISRO archived at the Indian Space Science Data Centre (ISSDC). This work has been performed utilising the calibration databases and auxiliary analysis tools developed, maintained and distributed by {\it AstroSat-SXT} team with members from various institutions in India and abroad. This research has made use of {\it MAXI} data provided by RIKEN, JAXA and the {\it MAXI} team. Also this research  {has made} use of software provided by the High Energy Astrophysics Science Archive Research Center (HEASARC) and NASA’s Astrophysics Data System Bibliographic Services. AN also thank GH, SAG; DD, PDMSA and Director-URSC for encouragement and continuous support to carry out this research.
\par
\noindent{Facilities: {\it AstroSat}, {\it MAXI},  {{\it NICER}}, {\it NuSTAR} .}
\section*{Data Availability}

The data used for analysis in this article are available in {\it AstroSat}-ISSDC website (\url{https://astrobrowse.issdc.gov.in/astro_archive/archive/Home.jsp}), {\it MAXI} website (\url{http://maxi.riken.jp/top/index.html}) and {\it NICER} and \textit{NuSTAR} observations from HEASARC database (\url{https://heasarc.gsfc.nasa.gov/docs/cgro/db-perl/W3Browse/w3browse.pl}).

\begin{landscape}
\begin{table}

\centering
 \caption{Spectral parameter values for GRS~1915$+$105 obtained from the best-fit model for each segment corresponding to the Re-brightening phases (see Table \ref{tab 2}). N${_{\rm H}}$ ($\times 10^{22}$  atoms cm$^{-2}$) corresponds to the column density due to the ISM absorption.  {kT$_\textrm{in}$} (keV) corresponds to \textit{diskbb} temperature,  { {F$_{\rm dbb}\times 10^{-9}$} is the unabsorbed disc flux in $0.3 - 100$ keV energy range (in erg s$^{-1}$cm$^{-2}$)}. $\Gamma$ denotes the Photon Index, $kT_\textrm{e}$ (keV) corresponds to electron temperature, and  { {F$_{\rm nth}\times 10^{-9}$} is the unabsorbed Comptonized flux in $0.3 - 100$ keV energy range (in erg s$^{-1}$cm$^{-2}$)}.  {E$_{\rm line}$ and EW components corresponding to the Gaussian model represent the central line energies and the equivalent width of the emission lines (in keV). Similarly the E$_{\rm Gabs}$ and the EW indicate the central energy and the equivalent width of the absorption lines (in keV).} \textit{TBpcf} is the partial covering absorption model along the continuum. The N${_{\rm H_{1}}}$ ($\times 10^{22}$  atoms cm$^{-2}$) under \textit{TBpcf} model indicates the local absorption value, the PCF denotes the local partial covering fraction.  Errors for all the parameters are calculated with 90\% confidence.}

\hspace*{-0.3cm}
\begin{tabular}{|lccccccccccccccc|}
\hline

 & TBabs & \multicolumn{2}{c}{TBpcf}   & \multicolumn{2}{c}{ {diskbb}} & \multicolumn{3}{c}{nthComp} & \multicolumn{2}{c}{Gauss} & \multicolumn{2}{c}{Gauss} & \multicolumn{2}{c}{Gabs} &   \\
\cline{2-15}
\rule{0pt}{10pt} \multirow{-2}{*}{Obs} &N${_{\rm H}\times10^{22}}$ & N${_{\rm H_{1}}\times10^{22}}$& PCF &  {kT$_{\rm in}$} &  {F$_{\rm dbb}\times 10^{-9}$} & $\Gamma$  & kT$_{\rm e}$ &  {F$_{\rm nth}\times 10^{-9}$} & E$_{\rm line}$& EW &E$_{\rm line}$& EW  & E$_{\rm Gabs}$&  {EW} & \multirow{-2}{*}{$\chi^{2}$/d.o.f} \\
 & (atm cm$^{-2}$) & (atm cm$^{-2}$) & &  {(keV)} &  {(erg s$^{-1}$cm$^{-2}$)} & &(keV) &  {(erg s$^{-1}$cm$^{-2}$)} & (keV) &  {(keV)} & (keV) &  {(keV)} & (keV) &  {(keV)} &  \\
\hline
\multicolumn{16}{|c|}{RB$_{\rm I}$} \\
\hline
Obs. 1 &  {5.04$_{-0.26}^{+0.11}$} &  {82.45$_{-4.12}^{+3.89}$} &  {0.70$_{-0.01}^{+0.02}$} &  {0.32*} &  {0.10$_{-0.03}^{+0.05}$} & {1.47$_{-0.02}^{+0.04}$} &  {2.65$_{-0.41}^{+0.19}$} &    {{0.28$_{-0.04}^{+0.07}$}} &   {6.38$_{-0.05}^{+0.09}$} &  {0.19$_{-0.01}^{+0.03}$} &  {6.67$_{-0.06}^{+0.10}$} &  {0.09$_{-0.01}^{+0.02}$} &  {-} &  {-} &   {128.42/129} \\

Obs. 2 &  {5.08$_{-0.19}^{+0.18}$} &  {33.73$_{-4.19}^{+4.00}$} &  {0.77$_{-0.01}^{+0.01}$} & - & - &  {1.29$_{-0.04}^{+0.02}$} &  {4.51$_{-0.52}^{+0.44}$} &   {{0.84$_{-0.09}^{+0.11}$}} &  {6.38$_{-0.18}^{+0.22}$} &  {0.08$_{-0.02}^{+0.02}$} &  {6.97$_{-0.26}^{+0.12}$} &  {0.12$_{-0.04}^{+0.04}$} &  {-} &  {-} &   {172.9/127} \\

 {Obs. 3} &  {5.17$_{-0.16}^{+0.20}$} &  {18.38$_{-0.76}^{+2.33}$} &  {0.69$\dagger$} & - & - &  {1.31$_{-0.02}^{+0.02}$} &  {4.32$_{-0.30}^{+0.38}$} &   {{4.39$_{-0.97}^{+1.30}$}} &  {6.39$_{-0.42}^{+0.29}$} &  {0.11$_{-0.01}^{+0.01}$} &  {6.95$_{-0.26}^{+0.37}$} &  {0.12$_{-0.01}^{+0.02}$} &  {-} &  {-} &   {178.4/133} \\

 {Obs. 4} &  {5.26$_{-0.59}^{+0.27}$} &  {50.95$_{-2.67}^{+2.32}$} &  {0.56$\dagger$} & - & - &  {1.43$_{-0.04}^{+0.03}$} &  {2.87$_{-0.51}^{+0.48}$} &   {{1.32$_{-0.09}^{+0.27}$}} &  {6.39$_{-0.26}^{+0.33}$} &  {0.13$_{-0.03}^{+0.04}$} &  {6.95$_{-0.19}^{+0.25}$} &  {0.14$_{-0.03}^{+0.04}$} &  {-} &  {-} &   {107.2/101} \\

 {Obs. 5} &  {5.55$_{-0.30}^{+0.22}$} &  {108.75$_{-4.69}^{+7.77}$} &  {0.72$\dagger$} & - & - &  {1.30$_{-0.01}^{+0.01}$} &  {1.63$_{-0.11}^{+0.27}$} &   {{0.06$_{-0.01}^{+0.01}$}} &  {6.37$_{-0.63}^{+0.21}$} &  {0.09$_{-0.01}^{+0.03}$} &  {6.65$_{-0.32}^{+0.40}$} &  {0.15$_{-0.02}^{+0.03}$} &  {-} &  {-} &  {110.52/92} \\
\hline
\multicolumn{16}{|c|}{RB$_{\rm II}$} \\
\hline
 {Obs. 1} &  {5.57$_{-0.03}^{+0.03}$} &   {16.42$_{-2.19}^{+3.44}$} &  {0.63$_{-0.02}^{+0.03}$} & - & - &  {1.20$_{-0.03}^{+0.02}$} &  {2.02$_{-0.16}^{+0.29}$} &   {1.16$_{-0.12}^{+0.09}$} & { 7.11$_{-0.14}^{+0.09}$ }&  {1.09$_{-0.14}^{+.19}$} &  {6.37$_{-0.37}^{+0.62}$} &  {0.44$\dagger$} &  {-} &  {-} &   {179/146}\\
\hline
\multicolumn{16}{|c|}{RB$_{\rm IV}$} \\
\hline
 {Obs. 1} &  {5.62$_{-0.06}^{+0.05}$} &  {78.08$_{-7.09}^{+4.27}$} &  {0.71$_{-0.02}^{+0.02}$} & - & - &  {1.16$_{-0.06}^{+0.06}$} &  {2.18$_{-0.13}^{+0.10}$} &   {0.97$_{-0.09}^{+0.09}$} &  {-} &  {-} &  {6.68$_{-0.21}^{+0.28}$} &  {1.98$_{-0.29}^{+0.33}$} &  {7.05$_{-0.10}^{+.34}$} &  {0.18$\dagger$} &   {137/143}\\
\hline
\multicolumn{16}{|c|}{RB$_{\rm V}$} \\
\hline
 {Obs. 1} &  {4.99$_{-0.21}^{+0.21}$} &   {150.56$_{-10.32}^{+12.77}$} &  {0.73$_{-0.02}^{+0.02}$} & - & - &   {1.34$_{-0.04}^{+0.03}$} &  {2.66$_{-0.26}^{+0.18}$} &   {0.70$_{-0.08}^{+0.08}$} &  {6.62$_{-0.22}^{+0.31}$}   &  { 0.15$_{-0.03}^{+0.04}$} &  {6.96$_{-0.14}^{+0.09}$} &  {0.10$_{-0.02}^{+0.04}$} &  {-} &  {-} &   {148.6/117}\\

 {Obs. 2} &  {5.43$_{-0.28}^{+0.39}$} &  {6.62$_{-0.66}^{+0.31}$} &  {0.60$\dagger$} & - & - &  {2.02$_{-0.06}^{+0.06}$} &  {1.85$_{-0.06}^{+0.20}$} &   {1.99$_{-0.90}^{+0.28}$} &  {6.67$_{-0.11}^{+0.26}$} &  {0.08$\dagger$} &  {6.37$_{-0.48}^{+0.63}$} &  {0.03$\dagger$} &  {-} &  {-}&   {131.5/126} \\

 {Obs. 3} &  {5.74$_{-0.41}^{+0.40}$} &  {8.52$_{-0.09}^{+0.09}$} &  {0.79$_{-0.03}^{+0.03}$} & - & - &  {2.47$_{-0.06}^{+0.03}$} &  {2.04$_{-0.34}^{+0.51}$} &   {2.21$_{-0.27}^{+0.36}$} &  {6.95$_{-0.17}^{+0.26}$} &  {0.07$\dagger$} &  {-}  &  {-} &  {6.53$_{-0.25}^{+0.37}$} &  {0.31$\dagger$} &   {166.8/148} \\

 {Obs. 4} &  {5.55$_{-0.27}^{+0.19}$} &  {4.37$_{-0.48}^{+0.31}$} &  {0.74$_{-0.01}^{+0.01}$} & - & - &  {1.91$_{-0.01}^{+0.03}$} &  {1.82$_{-0.39}^{+0.45}$} &   {0.40$_{-0.07}^{+0.04}$} &  {6.95$_{-0.47}^{+0.63}$} &  {0.12$_{-0.01}^{+0.01}$} &  {6.36$_{-0.38}^{+0.43}$} &  {0.12$_{-0.01}^{+0.02}$} &  {-} &  {-} &   {201.142/142} \\
\hline
\multicolumn{16}{|c|}{RB$_{\rm VI}$} \\
\hline
 {Obs. 1} &  {6.09$_{-0.65}^{+0.49}$} &   {98.48$_{-10.82}^{+4.79}$} &  {0.77$_{-0.02}^{+0.02}$} & - & - &  {1.36$_{-0.02}^{+0.02}$}  &  {1.94$_{-0.05}^{+0.11}$} &   {0.71$_{-0.09}^{+0.09}$} &  {-} &   {-} &  {-}    &   {-}    &  {6.56$_{-0.24}^{+0.21}$} &  {0.59$_{-0.11}^{+0.19}$} &   {162.8/127} \\

 {Obs. 2} &  {4.93$_{-0.33}^{+0.41}$} &  {4.42$_{-0.41}^{+0.63}$} &  {0.93$_{-0.01}^{+0.02}$} & - & - &  {1.55$_{-0.02}^{+0.02}$} &  {1.65$_{-0.38}^{+0.092}$} &   {4.29$_{-0.81}^{+1.21}$} &  {-}  &   {-}    &  {-} &   {-}  &  {6.65$_{-0.17}^{+0.10}$} &   {0.12$_{-0.03}^{+0.02}$} &   {138/136} \\

 {Obs. 3} &  {6.10$_{-0.35}^{+0.37}$} &  {6.30$_{-0.26}^{+0.67}$} &  {0.73$_{-0.03}^{+0.02}$} & - & - &  {2.04$_{-0.04}^{+0.02}$} &  {1.69$_{-0.28}^{+0.07}$} &   {4.90$_{-0.42}^{+0.61}$} &  {-}    &   {- }   &  {-}    &   {-}    &  {6.68$_{-0.12}^{+0.12}$} &  {0.19$_{-0.01}^{+0.02}$} &   {127.1/133} \\

 {Obs. 4} &  {6.22$_{-0.29}^{+0.35}$} &  {8.93$_{-0.61}^{+0.60}$} &  {0.37$\dagger$} & - & - &  {1.69$_{-0.05}^{+0.02}$} &  {1.74$_{-0.26}^{+0.34}$} &   {4.27$_{-0.73}^{+0.69}$} &  {-}    &  { -  }  &  {-}    &   {-}    &  {6.61$_{-0.13}^{+0.11}$} &  {0.23$_{-0.05}^{+0.05}$} &   {110.3/133} \\

 {Obs. 5} &  {4.82$_{-0.43}^{+0.41}$} &  {44.61$_{-3.60}^{+5.92}$} &  {0.95$\dagger$} & - & - &  {1.37$_{-0.04}^{+0.04}$} &  {2.19$_{-0.32}^{+0.22}$} &   {0.29$_{-0.04}^{+0.03}$} &  {-}    &   {-}    &  {-}    &   {-}    &  {6.58$_{-0.26}^{+0.29}$} &  {0.39$_{-0.01}^{+0.01}$} &   {144.97/106} \\
\hline
\end{tabular}
\label{tab 4}
\begin{flushleft}
      * Frozen ;  $\dagger$ Error values are insignificant 
\end{flushleft}
\end{table}
\end{landscape}

\begin{landscape}
\begin{table}
\centering
 \caption{ Spectral parameter values for GRS~1915$+$105 obtained from the best-fit model for each Epoch corresponding to the wide-band observations (see Table \ref{tab 1}).   N${_{\rm H}}$ ($\times 10^{22}$  atoms cm$^{-2}$) corresponds to the column density due to the ISM absorption.  {kT$_\textrm{in}$} (keV) corresponds to \textit{diskbb} temperature,  { {F$_{\rm dbb}\times 10^{-9}$} is the unabsorbed disc flux in $0.3 - 100$ keV energy range (in erg s$^{-1}$cm$^{-2}$)}. $\Gamma$ denotes the Photon Index, $kT_\textrm{e}$ (keV) corresponds to electron temperature, and  { {F$_{\rm nth}\times 10^{-9}$} is the unabsorbed Comptonized flux in $0.3 - 100$ keV energy range (in erg s$^{-1}$cm$^{-2}$)}.  {E$_{\rm line}$ and EW components under the \textit{gaussian} model represent the emission line energies and the equivalent width of the line.  E$_{\rm Gabs}$ and EW correspond to absorption line and the equivalent width of the absorption line (in keV).} E$_{\rm Smed}$ and  Width parameters under the \textit{Smedge} component corresponds to the energy and the width of the smeared edge component (in keV).  \textit{TBpcf} is the partial covering absorption model along the continuum. The N${_{\rm H_{1}}}$ ($\times 10^{22}$  atoms cm$^{-2}$) under \textit{TBpcf} model indicates the local absorption value, the PCF denotes the local partial covering fraction.  Errors for all the parameters are calculated with 90\% confidence.}
 \label{tab 6}
 \hspace*{-0.55cm}
\begin{tabular}{|lccccccccccccccc|}
\hline
&TBabs & \multicolumn{2}{c}{TBpcf} & \multicolumn{2}{c}{diskbb} & \multicolumn{3}{c}{nthComp} & \multicolumn{2}{c}{Gauss} & \multicolumn{2}{c}{Gabs} & \multicolumn{2}{c}{Smedge} & \\
\cline{2-15}  
\rule{0pt}{10pt} \multirow{-2}{*}{Ep.} &N${_{\rm H}\times10^{22}}$ & N${_{\rm H_{1}}\times10^{22}}$& PCF &kT$_{\rm in}$&  {F$_{\rm dbb}\times 10^{-9}$} &$\Gamma$& kT$_{\rm e}$&  {F$_{\rm nth}\times 10^{-9}$}  & E$_{\rm line}$&  {EW}  & E$_{\rm Gabs}$&  {EW} & E$_{\rm Smed}$& Width & \multirow{-2}{*}{$\chi^{2}$/d.o.f} \\
 & (atm cm$^{-2}$) & (atm cm$^{-2}$) & & (keV) &  {(erg s$^{-1}$cm$^{-2}$)} & & (keV) &  {(erg s$^{-1}$cm$^{-2}$)} & (keV) &  {(keV)} & (keV) &  {(keV)} & (keV) & (keV) &  \\
\hline
 1 & 5.13$_{-0.21}^{+0.23}$ & - & - & 1.36$_{-0.02}^{+0.02}$ &  {1.19$_{-0.59}^{+0.73}$} & 1.60$_{-0.03}^{+0.03}$ & 11.6$_{-0.4}^{+0.4}$ &  {3.02$_{-1.04}^{+1.28}$} & 6.88$_{-0.15}^{+0.18}$  &  {0.99$_{-0.09}^{+0.07}$} & - & - & - & - & 445.56/473\\
 
 2 & 4.16$_{-0.29}^{+0.30}$ & 30.68$_{-2.17}^{+2.27}$ & 0.70$_{-0.02}^{+0.02}$ & 1.16$_{-0.02}^{+0.02}$ &  {1.06$_{-0.10}^{+0.52}$} & 1.73$_{-0.02}^{+0.02}$ & 14.97$_{-1.72}^{+2.53}$ &  {1.58$_{-0.61}^{+0.79}$} & -  & - & - & - & 7.41$_{-0.13}^{+0.13}$ & 1.08$_{-0.35}^{+0.48}$ & 472.44/496\\

 3 & 3.61$_{-0.11}^{+0.13}$ & 45.2$_{-4.6}^{+8.1}$ & 0.13$_{-0.02}^{+0.02}$ & 1.32$_{-0.04}^{+0.05}$ &  {0.72$_{-0.09}^{+0.18}$} & 1.13$_{-0.06}^{+0.06}$ & 9.5$_{-0.8}^{+1.2}$ &  {0.91$_{-0.09}^{+0.15}$} & 6.47$_{-0.14}^{+0.21}$  &  {1.20$_{-0.11}^{+0.17}$} & - & - & - & - & 79.61/87\\
 
 4 & 7.35$_{-0.32}^{+0.35}$ & 612.28$_{-72.60}^{+85.62}$ & 0.42$_{-0.01}^{+0.01}$ & 0.24$_{-0.01}^{+0.02}$ &  {0.23$_{-0.06}^{+0.07}$} & 1.28$_{-0.03}^{+0.04}$ & 8.22$_{-0.16}^{+0.16}$ &  {0.55$_{-0.07}^{+0.06}$} & 6.54$_{-0.02}^{+0.02}$  &  {1.69$_{-0.83}^{+0.53}$} & - & - & - & - & 478.59/479\\
 
 5 & 6.02$_{-0.20}^{+0.22}$ & 108.0$_{-18.3}^{+11.7}$ & 0.41$_{-0.01}^{+0.01}$ & 1.09$_{-0.07}^{+0.09}$ &  {0.40$_{-0.07}^{+0.04}$}  & 1.21$_{-0.05}^{+0.06}$ & 16.6$_{-1.4}^{+1.7}$ &  {0.52$_{-0.06}^{+0.06}$} & 6.63$_{-0.31}^{+0.29}$  &  {1.09$_{-0.10}^{+0.10}$} & - & - & - & - & 139.7/134\\
 
 6 & 5.39$_{-0.32}^{+0.35}$ & 544.61$_{-72.60}^{+85.62}$ & 0.38$_{-0.01}^{+0.01}$ & - & - & 1.28$_{-0.06}^{+0.04}$ & 6.74$_{-0.16}^{+0.16}$ &  {1.17$_{-0.11}^{+0.47}$}  & 6.53$_{-0.02}^{+0.02}$  &  {1.89$_{0.21}^{+0.24}$} & - & - & 8.70$_{-0.20}^{+0.17}$ & 0.47$_{-0.28}^{+0.38}$ & 566.66/529\\
 
 7 & 6.36$_{-0.47}^{+0.48}$ & 24.67$_{-2.18}^{+2.64}$ & 0.64$_{-0.04}^{+0.04}$ & - & - & 1.40$\dagger$ & 6.37$_{-0.07}^{+0.16}$ &  {1.61$_{-0.80}^{+0.95}$}  & 6.55$_{-0.03}^{+0.02}$  &  {2.32$_{-0.66}^{+0.61}$} & - & - & 8.87$_{-0.22}^{+0.18}$ & 1.74$_{-0.38}^{+0.47}$ & 866.2/786\\
 
 8 & 5* & 29.79$_{-2.73}^{+2.91}$ & 0.48$_{-0.04}^{+0.03}$ & - & - & 1.30$_{-0.02}^{+0.02}$ & 7.47$_{-0.20}^{+0.40}$ &  {1.04$_{-0.62}^{+0.39}$}  & 6.53$_{-0.03}^{+0.03}$ &  {1.19$_{-0.12}^{+0.10}$} & - & - & 7.57$_{-0.39}^{+0.35}$ & 3.21$_{-0.32}^{+0.35}$ & 382.44/388\\
 
 9 & 5.24$_{-0.56}^{+0.58}$ & 35.41$_{-3.38}^{+3.87}$ & 0.73$_{-0.04}^{+0.03}$ & - & - & 1.22$\dagger$ & 8.67$_{-0.41}^{+0.48}$ &  {0.64$_{-0.08}^{+0.09}$}  & 6.55$_{-0.03}^{+0.03}$ &  {1.31$_{-0.09}^{+0.11}$} & - & - & - & - & 552.83/452\\

 10 & 6.35$_{-0.81}^{+0.85}$ & 22.70$_{-1.66}^{+2.08}$ & 0.66$_{-0.08}^{+0.06}$ & 1.12$_{-0.04}^{+0.04}$ &  {1.77$_{-0.90}^{+0.83}$}  & 1.85$_{-0.06}^{+0.06}$ & 12.86$_{-0.44}^{+0.50}$ &  {0.73$_{-0.17}^{+0.09}$}  & 6.85$_{-0.28}^{+0.25}$  &  {3.26$_{-0.75}^{+0.62}$} & - & - & - & - & 655.40/607\\
 
 11 & 5* & 16.15$_{-0.96}^{+0.99}$ & 0.56$_{-0.08}^{+0.06}$ & - & - & 1.24$_{-0.01}^{+0.02}$ & 8.19$_{-0.41}^{+0.44}$ &  {0.43$_{-0.04}^{+0.04}$}  & 6.52$_{-0.05}^{+0.05}$  &  {1.19$_{-0.02}^{+0.21}$} & - &- & 8.54$_{-0.51}^{+0.73}$ & 1.27$_{-0.37}^{+0.72}$& 131.61/137\\
 
 12 & 5* & 52.11$_{-2.01}^{+2.01}$ & 0.76$\dagger$ & - & - & 1.65$_{-0.018}^{+0.02}$ & 6.67$_{-0.26}^{+0.29}$ &  {1.99$_{-0.74}^{+0.46}$}  & 6.48$_{-0.03}^{+0.03}$  &  {1.00$_{-0.19}^{+0.08}$} & - & - & 8.02$_{-0.68}^{+0.59}$ & 4.17$_{-0.47}^{+0.46}$ & 527.02/509\\
 
 13 & 6.44$_{-0.04}^{+0.04}$ & 27.13$_{-0.91}^{+0.94}$ & 0.28$\dagger$ & 1.59$_{-0.02}^{+0.01}$ &  {1.85$_{-0.64}^{+0.55}$}  & 2.76$_{-0.04}^{+0.04}$ & 3.09$_{-0.14}^{+0.15}$ &  {1.79$_{-0.19}^{+0.38}$}  & -  & - & 6.66$_{-0.01}^{+0.01}$ &  {0.12$_{-0.01}^{+0.01}$} & 8.52$_{-0.07}^{+0.07}$ & 0.32$_{-0.09}^{+0.08}$ & 1186.51/1094\\
 
 14 & 6.01$_{-0.05}^{+0.06}$ & 18.48$_{-0.76}^{+0.77}$ & 0.73$\dagger$ & 1.0$_{-0.02}^{+0.02}$ &  {1.48$_{-0.09}^{+0.13}$}  & 2.21$_{-0.24}^{+0.23}$ & 3.49$_{-0.27}^{+0.35}$ &  {1.53$_{-0.17}^{+0.67}$}  & 7.34$_{-0.23}^{+0.15}$  &  {2.39$_{-0.20}^{+0.19}$} & - & - & - & - & 491.66/511\\
 
 15 & 6.25$_{-0.09}^{+0.09}$ & 27.61$_{-0.42}^{+0.43}$ & 0.81$\dagger$ & 0.99$\dagger$ &  {1.18$_{-0.06}^{+0.08}$}  & 2.30$_{-0.31}^{+0.28}$ & 3.90$_{-0.29}^{+0.33}$ &  {1.57$_{-0.45}^{+0.92}$}  & 7.47$_{-0.20}^{+0.22}$ &  {2.91$_{-0.19}^{+0.25}$} & - & - & - & - & 693.97/624\\
 
 16 & 6.35$_{-0.15}^{+0.15}$ & 24.26$_{-0.76}^{+0.77}$ & 0.81$\dagger$ & 0.97$\dagger$ &  {1.23$_{-0.19}^{+0.23}$}  & 2.69$_{-0.05}^{+0.05}$ & 4.94$_{-0.30}^{+0.34}$ &  {1.49$_{-0.27}^{+0.71}$}  & 7.42$_{-0.13}^{+0.15}$ &  {2.67$_{-0.30}^{+0.41}$} & - & - & - & - & 585.13/554\\
 
 17 & 4.11$_{-0.07}^{+0.08}$ & 21.26$_{-1.04}^{+1.06}$ & 0.88$\dagger$ & - & - & 1.71$_{-0.02}^{+0.01}$ & 31.40$_{-1.49}^{+1.56}$ &  {0.78$_{-0.06}^{+0.06}$}  & - & - & 6.57$_{-1.23}^{+1.20}$ &  {0.29$_{-0.01}^{+0.01}$} & - & - & 755.44/800\\
 
 18 & 4.41$_{-0.75}^{+0.84}$ & 101.01$_{-3.08}^{+3.22}$ & 0.96$\dagger$ & - & - & 1.47$_{-0.01}^{+0.02}$ & 18.17$_{-3.71}^{+4.81}$ &  {0.80$_{-0.04}^{+0.09}$}  & 6.14$_{-0.18}^{+0.12}$ &  {0.99$_{-0.09}^{+0.07}$} & - & - & - & - & 234.78/254\\
\hline
\end{tabular}
\begin{flushleft}
      * Frozen ;  $\dagger$ Error values are insignificant 
\end{flushleft}
\end{table}
\end{landscape}








\bibliographystyle{mnras}
\bibliography{reference} 

\bsp  
\label{lastpage}
\end{document}